\documentclass[11pt]{amsart}
\usepackage{ifthen,amsmath,amssymb,latexsym,epsf,amsthm,amscd,times}
\usepackage[nobysame]{amsrefs}
\newboolean{draft}
\setboolean{draft}{false}
\ifthenelse{\boolean{draft}}{\usepackage{showkeys}}{}

\newtheoremstyle{SmallCaps}{}{}{\itshape}{}{\textsc\bgroup}{.\egroup}{
 }{}

\newtheorem{dfn}{Definition}[section]

\newtheorem{pro}{Proposition}[section]

\newtheorem{thm}{Theorem}[section]
\newcommand{\dref}[1]{Definition~\ref{#1}}
\newcommand{\cref}[1]{Corollary~\ref{#1}}

\newcommand{\pref}[1]{Proposition~\ref{#1}}

\newcommand{\tref}[1]{Theorem~\ref{#1}}

\def\R{\mathbb{R}}
\def\C{\mathbb{C}}
\def\rk{\operatorname{rk}}
\def\Q{\mathbb{Q}}
\def\Z{\mathbb{Z}}
\def\N{\mathbb{N}}

\def\P{\mathbb{P}}

\def\Lhi{\mathcal{L}}

\newcommand{\onatop}[2]{\genfrac{}{}{0pt}{}{#1}{#2}}

\def\gammaeiner{\;\raisebox{-0.8cm}{\epsfysize=2cm\epsfbox{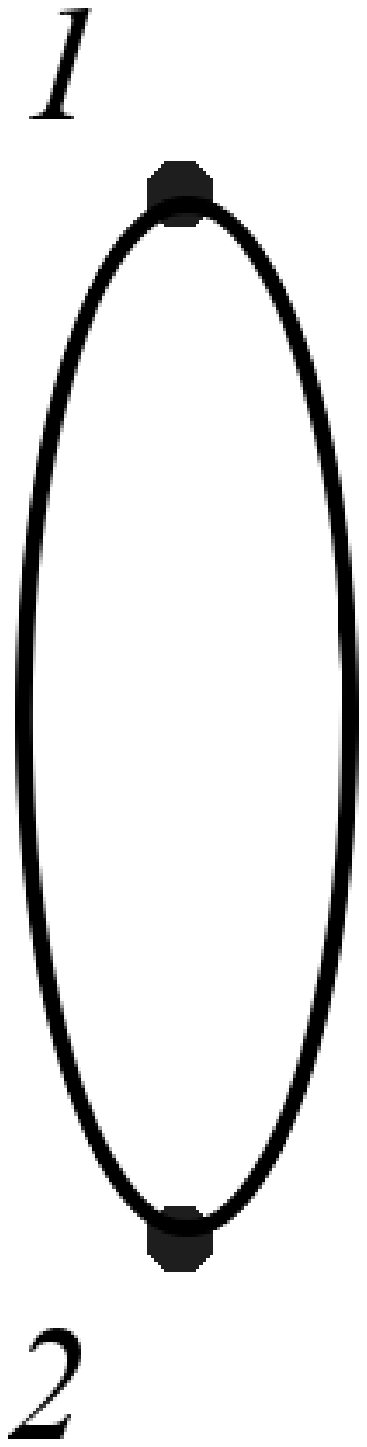}}\;}
\def\gammafive{\;\raisebox{-0.75cm}{\epsfysize=1.9cm\epsfbox{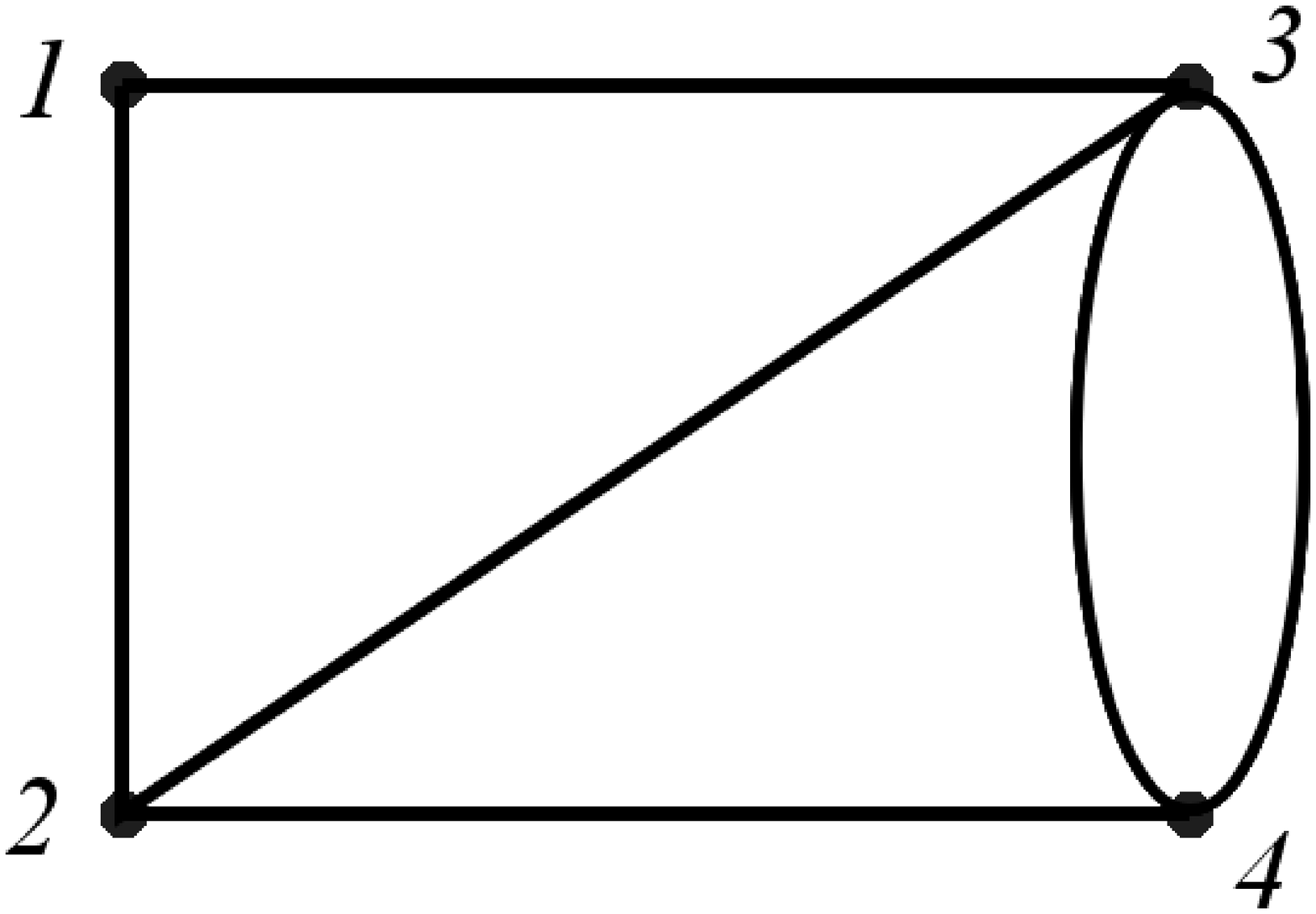}}\;}
\def\dunce{\;\raisebox{-0.75cm}{\epsfysize=1.9cm\epsfbox{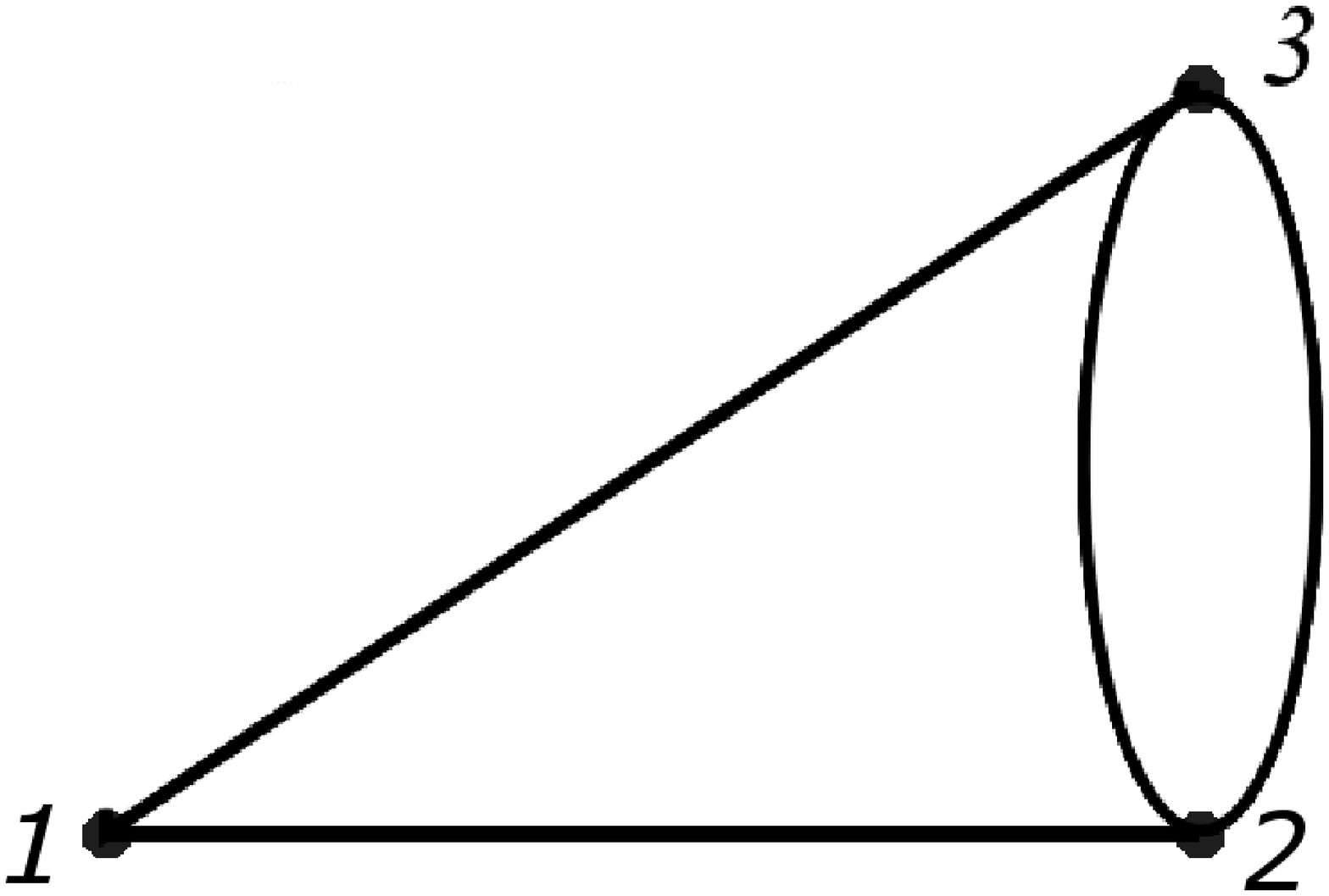}}\;}
\def\scfive{\;\raisebox{-0.6cm}{\epsfysize=1.5cm\epsfbox{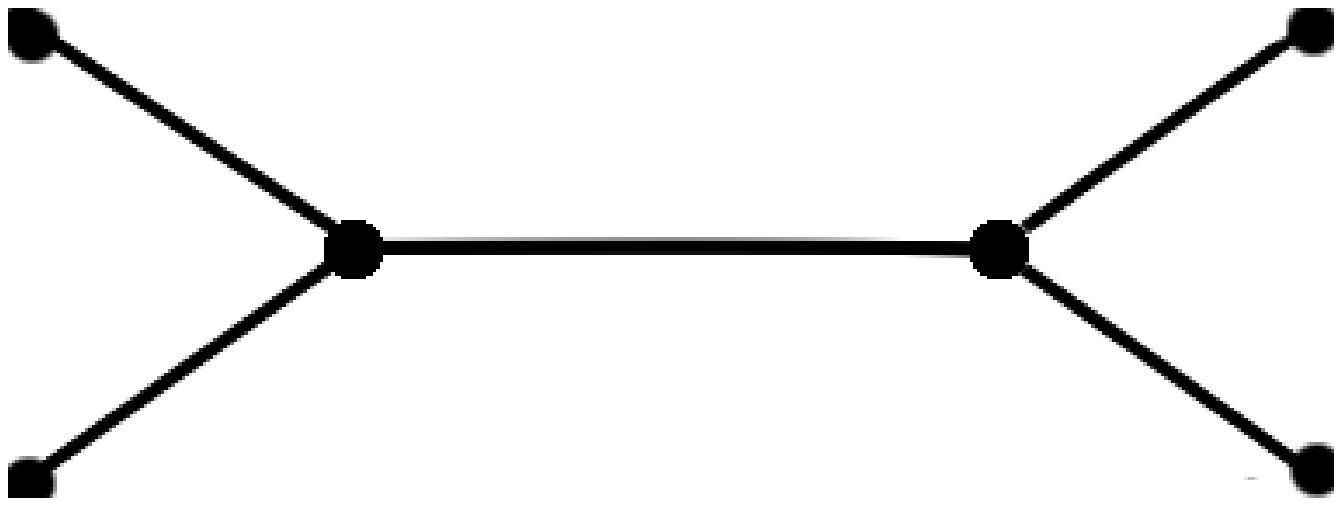}}\;}
\def\scsix{\;\raisebox{-0.6cm}{\epsfysize=1.5cm\epsfbox{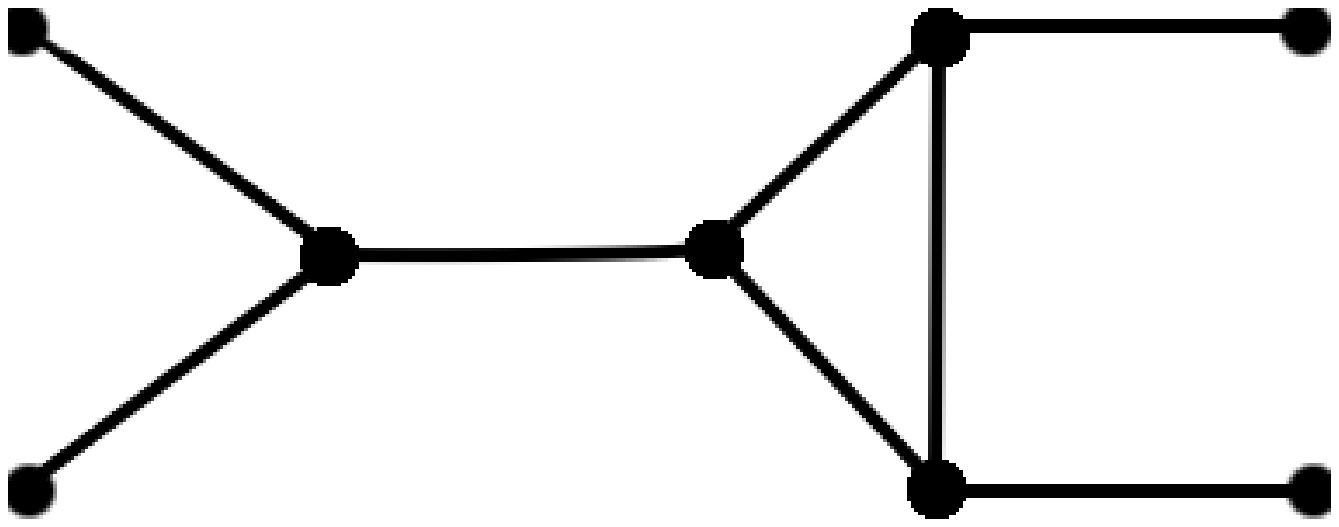}}\;}
\def\nest{\;\raisebox{-1.0cm}{\epsfysize=4.0cm\epsfbox{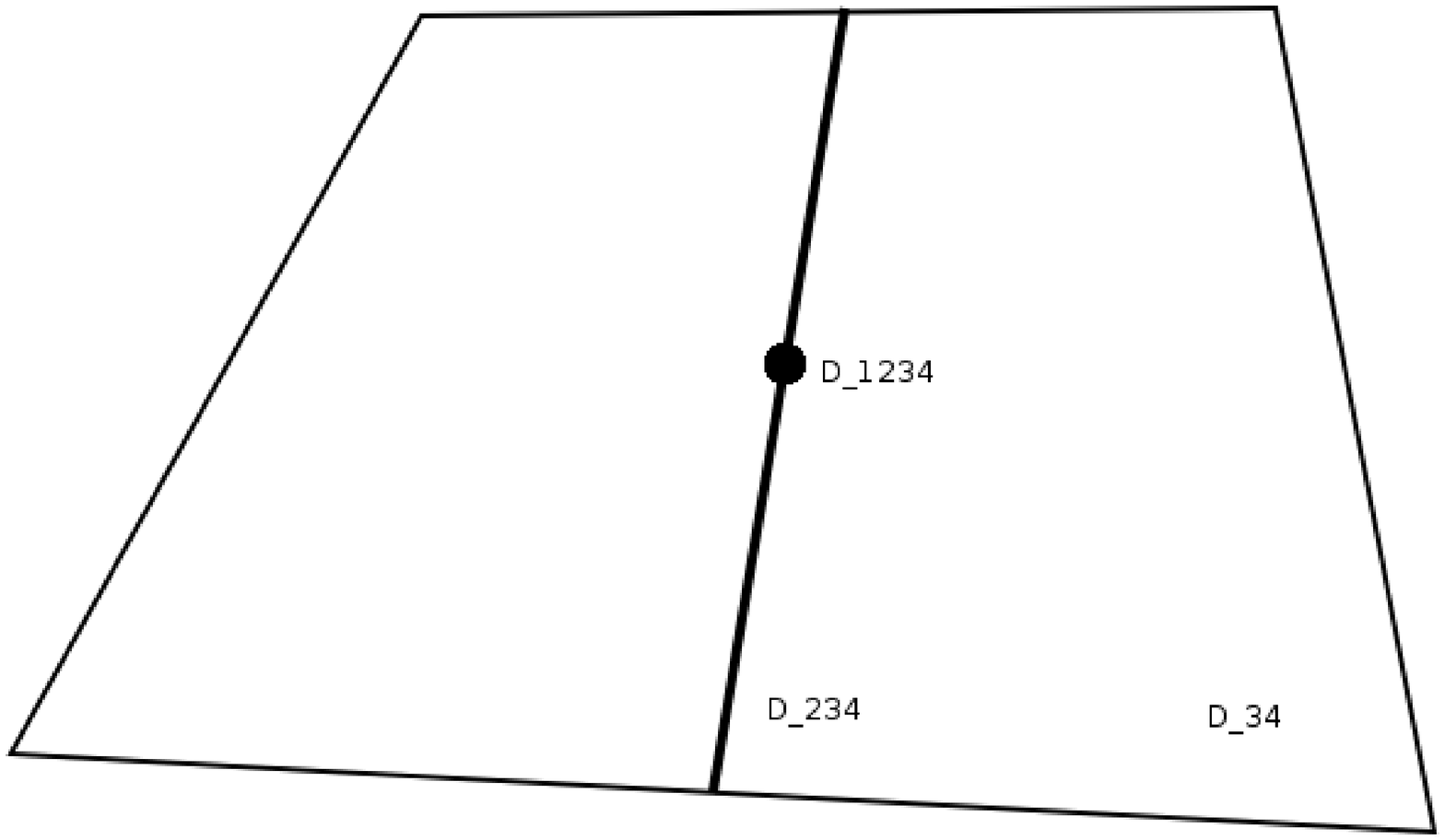}}\;}
\def\blowup{\;\raisebox{-1.0cm}{\epsfysize=4.0cm\epsfbox{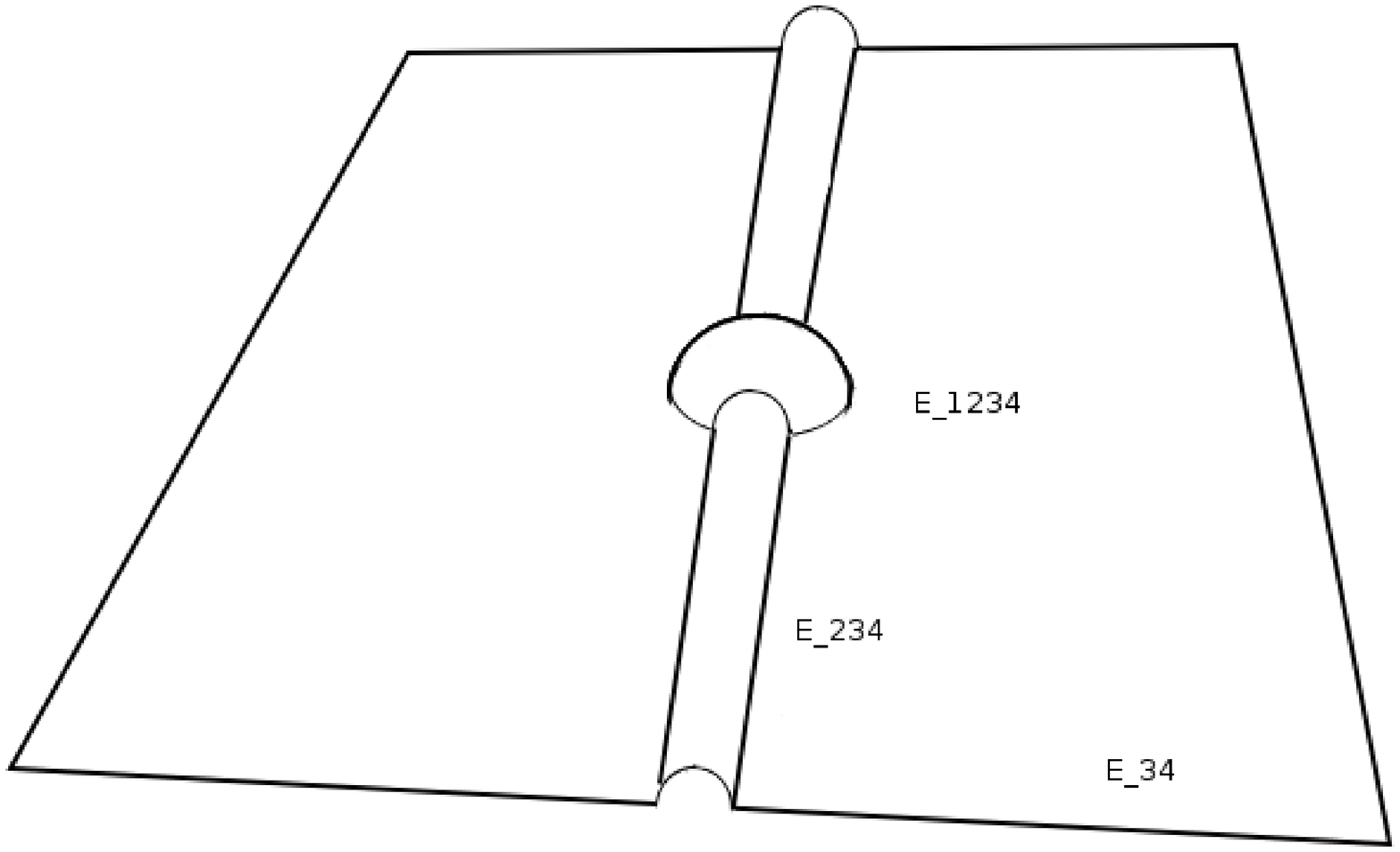}}\;}
\def\wheel{\;\raisebox{-0.6cm}{\epsfysize=1.5cm\epsfbox{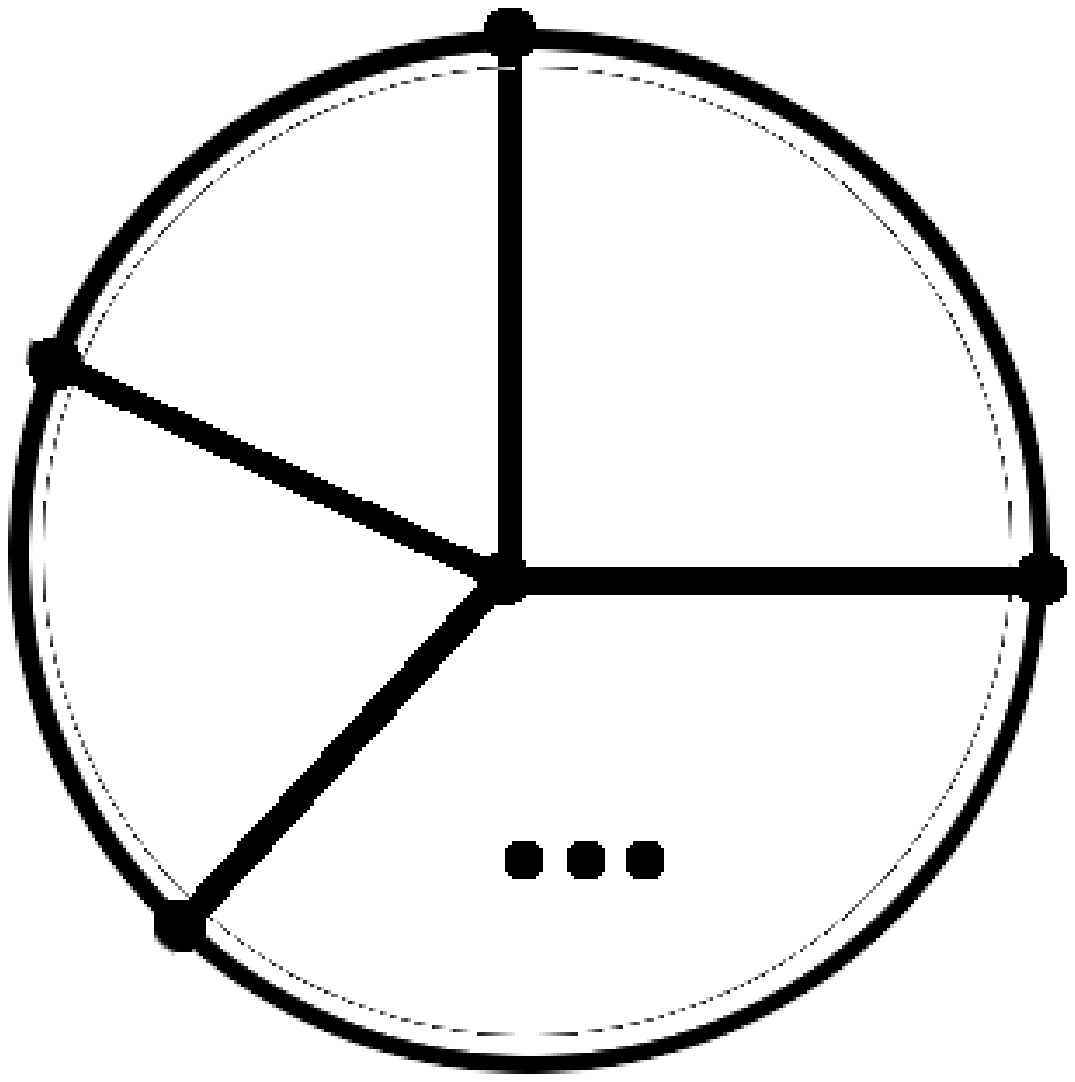}}\;}

\begin{document}
\title{Notes on Feynman Integrals and Renormalization}
\begin{abstract}
I review various aspects of Feynman integrals, regularization and renormalization.
Following Bloch, I focus on a linear algebraic approach to the Feynman rules, and I try to bring together
several renormalization methods found in the literature from a unifying point of view, using resolutions of singularities.
In the second part of the paper, I briefly sketch the work of Belkale, Brosnan resp.~ Bloch, Esnault and Kreimer
on the motivic nature of Feynman integrals.
\end{abstract}
\author{Christoph Bergbauer}

\keywords{Feynman graph, Feynman integral, Feynman rules,
regularization, renormalization, subspace arrangement, resolution of singularities, Connes-Kreimer Hopf algebra,
matroid, multiple zeta value, motive, period}

\address{SFB 45\\Institut f\"ur Mathematik\\Johannes-Gutenberg-Universit\"at Mainz}
\email{bergbau@math.fu-berlin.de}
\ifthenelse{\boolean{draft}}{\date{\today\, (\bf{DRAFT VERSION: PLEASE DO NOT CIRCULATE})}}{\date{May 20, 2010}}
\maketitle
\setcounter{tocdepth}{1} \tableofcontents
\section{Introduction}\label{s:intro}
In recent years there has been a growing interest in Feynman graphs and their integrals.

Physicists use Feynman graphs and
the associated integrals in order to compute certain experimentally measurable quantities out of quantum field theories. The problem
is that there are conceptual difficulties in the definition of interacting quantum field theories in four dimensions. The good thing is
that nonetheless the Feynman graph formalism is very successful in the sense that the quantities obtained from it match with the quantities obtained in
experiment extremely well. Feynman graphs are interpreted as elements of a perturbation theory, i.~e.~ as an expansion of an
(interesting) interacting quantum field theory in the neighborhood of a (simple) free quantum field theory. One therefore hopes
that a better understanding of Feynman graphs and their integrals could eventually lead to a better understanding of the true
nature of quantum field theories, and contribute to some of the longstanding open questions in the field. \\

A Feynman graph is simply a finite graph, to which one
associates a certain integral: The integrand depends on the
quantum field theory in question, but in the simplest case it
is just the inverse of a direct product of rank 4 quadratic
forms, one for each edge of the graph, restricted to a real
linear subspace determined by the topology of the graph.

For a general graph, there is currently no canonical way of
solving this integral analytically. However, in this simple
case where the integrand is algebraic, one can be convinced to
regard the integral as a period of a mixed motive, another
notion which is not rigorously defined as of today. All these
Feynman periods that have been computed so far, are rational
linear combinations of multiple zeta values, which are known to
be periods of mixed Tate motives, a simpler, and better
understood kind of motives. A stunning theorem of Belkale and
Brosnan however indicates that this is possibly a coincidence
due to the relatively small number of Feynman periods known
today: They showed that in fact any algebraic variety defined over $\Z$
is related to a Feynman graph hypersurface (the Feynman period
is one period of the motive of this hypersurface) in a quite obscure way. \\

The purpose of this paper is to review selected aspects of Feynman graphs, Feynman integrals and renormalization
in order to discuss some of the recent work by Bloch, Esnault, Kreimer and others on the motivic nature of these integrals.
It is based on public lectures given at the ESI in March 2009, at the DESY and IHES in April and June 2009, and several
informal lectures in a local seminar in Mainz in fall and winter 2009. I would like to thank the other participants for their
lectures and discussions.\\

Much
of my approach is centered around the notion of renormalization, which seems crucial for a deeper understanding of Quantum Field Theory.
No claim of originality is made except for section \ref{ss:ressing} and parts of the surrounding sections,
which is a review of my own research with R.~Brunetti and D.~Kreimer \cite{BBK}, and section \ref{ss:misc} which contains new results. \\

This paper is not meant to be a complete and up to date survey by any means.
In particular, several recent developments in the area, for example the work of Brown \cite{Brown2,Brown3}, Aluffi and Marcolli \cite{AM1,AM2,AM4}, Doryn and Schnetz \cite{Doryntalk,Schnetz},
 and the theory of Connes and Marcolli \cite{CoMaBook} are not covered here.
\subsubsection*{Acknowledgements} I thank S.~M\"uller-Stach, R.~Brunetti, S.~Bloch, M.~Kontsevich, P.~Brosnan, E.~Vogt, C.~Lange, A.~Usnich, T.~Ledwig, F.~Brown and especially D.~Kreimer for discussion on the subject of this paper. I would like to thank the ESI and the organizers of the spring 2009 program on number theory and physics for hospitality during the month of March 2009, and the IHES for hospitality in January and February 2010. My research is funded by the SFB 45 of the Deutsche Forschungsgemeinschaft.
\section{Feynman graphs and Feynman integrals}\label{s:feynman}
For the purpose of this paper, a Feynman graph is simply a finite connected multigraph where ''multi'' means that there may be
several, parallel edges between vertices. Loops, i.~e.~ edges connecting to the same vertex at both ends, are not allowed in this paper.
Roughly, physicists think of edges as virtual particles and of vertices as interactions between the virtual particles corresponding
to the adjacent edges.

If one has to consider several types of particles, one has several types (colors, shapes etc.) of edges.

Here is
an example of a Feynman graph:
\begin{equation*}
\scfive
\end{equation*}
This Feynman graph describes a theoretical process within a
scattering experiment: a pair of particles annihilates into a
third, intermediate, particle, and this third particle then
decays into the two outgoing particles at the right.

This Feynman graph (and the probability amplitude assigned to it)
make sense only as a single term in a first order approximation. In order to compute the scattering cross section, one will have
to sum over arbitrarily complicated Feynman graphs with four fixed external edges, and in this sum an infinity of graphs with cycles will
occur, for example
\begin{equation*}
\scsix
\end{equation*}
In this paper we will be concerned only with Feynman graphs
containing cycles, and I will simply omit the external edges
that correspond to the (asymptotic) incoming and outgoing
physical particles of a scattering experiment.

I will come back to the physical interpretation in greater detail in section
\ref{ss:physics}.
\subsection{Feynman rules}\label{ss:rules}
Feynman graphs are not only a nice tool for drawing complex interactions of virtual particles, they also provide a
recipe to compute the probability that certain scattering processes occur. The theoretical reason for this will be
explained later, but to state it very briefly, a Feynman graph is regarded as a \emph{label} for a term in a
perturbative expansion of this probability amplitude. This term in this expansion is called \emph{Feynman integral},
but at this point one must be careful with the word integral because of reasons of convergence.
\begin{dfn}\label{dfn:integral} An \emph{integral} is a pair $(A,u)$ where $A$ is an open subset of some $\R^n$ or $\R^n_{\ge 0},$ and $u$ a distribution in $A\cap \left(\R^n\setminus\bigcup H_i\right)$ where $H_i$ are affine subspaces.
\end{dfn}
A distribution in $X$ is a continuous linear functional on the space of \emph{compactly supported} test functions $C^\infty_0(X)$
with the usual topology. Locally integrable functions (that is, functions integrable on compact subsets) define distributions in an obvious way. Let us denote by $\underline{1}_{A}$ the characteristic function of $A$ in $\R^n.$ It is certainly not a test function unless $A$ is compact, but if $u$ allows (decays rapidly enough at $\infty$), then we may evaluate $u$ against $\underline{1}_A.$ We
write $u[f]$ for the distribution applied to the test function $f.$ If $u$ is given by a locally integrable function, we may also write $\int u(x) f(x) dx.$

If $u$ is given by a function which is integrable in all of $A,$ then $(A,u)$ can be associated with the usual integral $\int_A u(x)dx= u[\underline{1}_A].$ Feynman integrals
however are very often divergent: This means by definition that $\int_A u(x)dx$ is divergent, and this can either result from
problems with local integrability at the $H_i$ or lack of integrability at $\infty$ away from the $H_i$ (if $A$ is unbounded),
or both. (A more unified point of view would be to start with a $\P^n$ instead of $\R^n$ in order to have the divergence at
$\infty$ as a divergence at the hyperplane $H_\infty$ at $\infty,$ but I will not exploit this here). \\

A basic example for such a divergent integral is the pair $A=\R \setminus \{0\}$ and $u(x)=|x|^{-1}.$ The function $u$ is locally
integrable inside $A,$ hence a distribution in $A.$ But neither is it integrable as $|x|\rightarrow \infty,$ nor locally integrable at $\{0\}.$ We will see in a moment that the divergent Feynman integrals to be defined are higher-dimensional generalizations of this example, with an interesting arrangement of the $H_i.$ \\

The following approach, which I learned from S.~Bloch \cite{BEK,BlochJapan}, is quite powerful when one wants to understand
the various Feynman rules from a common point of view. It is based on the idea that a Feynman graph first defines a point configuration in some $\R^n,$ and it is only this point configuration which determines the Feynman integral via the Feynman
rules.\\

Let $\Gamma$ be a Feynman graph with set of edges $E(\Gamma)$ and set of vertices $V(\Gamma).$ A subgraph $\gamma$ has
by definition the same vertex set $V(\gamma)=V(\Gamma)$ but $E(\gamma)\subseteq E(\Gamma).$ Impose temporarily
an orientation of the edges, such that every edge has an incoming $v_{e,in}$ and an outgoing vertex $v_{e,out}.$ Since we do not allow loops, the two are different.
Set $(v:e)=1$ if $v$ is the outgoing
vertex of $e,$ $(v:e)=-1$ if $v$ is the incoming vertex and $e,$ and $(v:e)=0$ otherwise. Let $\mathcal{M}=\R^d,$ where
$d\in 2+2\N,$ called \emph{space-time}, with euclidean metric $|\cdot|.$ We will mostly consider the case where $d=4,$ but it is useful to see
the explicit dependence on $d$ in the formulas. \\

All the information of $\Gamma$ is encoded in the map
\begin{equation*}
\Z^{E(\Gamma)}\stackrel{\partial}{\rightarrow} \Z^{V(\Gamma)}
\end{equation*}
sending an edge $e\in E(\Gamma)$ to $\partial(e)=\sum_{v\in V(\Gamma)}(v:e)v=v_{e,out}-v_{e,in}.$ This is nothing but the
chain complex for the oriented simplicial homology of the 1-dimensional simplicial complex $\Gamma,$ and it is a standard construction to build from this map $\partial$ an exact sequence
\begin{equation}\label{eq:exactseq}
0\rightarrow H_1(\Gamma;\Z)\rightarrow \Z^{E(\Gamma)}\stackrel{\partial}{\rightarrow} \Z^{V(\Gamma)}\stackrel{}{\rightarrow} H_0(\Gamma;\Z) \rightarrow 0.
\end{equation}
Like this one obtains two inclusions of free abelian groups into $\Z^{E(\Gamma)}:$
\begin{equation*}
i_\Gamma: H_1(\Gamma;\Z)\hookrightarrow \Z^{E(\Gamma)}
\end{equation*}
The second one is obtained by dualizing
\begin{equation*}
j_\Gamma: \Z^{V(\Gamma)\vee}/H^0(\Gamma;\Z) \stackrel{\partial^\vee}{\hookrightarrow} \Z^{E(\Gamma)\vee}.
\end{equation*}
Here, and generally whenever a basis is fixed, we can canonically identify free abelian groups with their duals.

All this can be tensored with $\R,$ and we get inclusions $i_{\Gamma},$ $j_{\Gamma}$ of vector spaces into another vector space with a \emph{fixed basis.} If one then replaces any $\R^n$ by $\mathcal{M}^n$ and denotes $i_{\Gamma}^{\oplus d}=(i_{\Gamma},
\ldots,i_{\Gamma}),$ $j_{\Gamma}^{\oplus d}=(j_{\Gamma},\ldots,j_{\Gamma}),$ then two types of Feynman integrals $(A,u)$ are
defined as follows:
\begin{eqnarray*}
A_{M}=H_1(\Gamma;\R)^d, && u^{M}_{\Gamma}=(i_\Gamma^{\oplus d})^\ast u_{0,M}^{\otimes |E(\Gamma)|},\\
A_{P}=\mathcal{M}^{V(\Gamma)\vee}/H^0(\Gamma;\R)^d, && u^{P}_{\Gamma}=(j_\Gamma^{\oplus d})^\ast u_{0,P}^{\otimes|E(\Gamma)|}.
\end{eqnarray*}
The distributions $u_{0,M},u_{0,P}\in \mathcal{D}'(\mathcal{M})$ therein are called \emph{momentum space} resp. \emph{position space propagators}.
Several examples of propagators and how they are related will be discussed in the next section, but for a first
reading
\begin{equation*}
u_{0,M}(p)=\frac{1}{|p|^2}, \quad u_{0,P}(x)=\frac{1}{|x|^{d-2}},
\end{equation*}
inverse powers of a rank $d$ quadratic form. As announced earlier, the pullbacks $(i_\Gamma^{\oplus d})^\ast u_{0,M}^{\otimes |E(\Gamma)|}$ and
$(j_\Gamma^{\oplus d})^\ast u_{0,M}^{\otimes |E(\Gamma)|}$ are only defined as distributions outside certain affine spaces
$H_i,$ that is for test functions supported on compact subsets which do not meet these $H_i.$ \\

The map
\begin{equation*}
\Gamma\mapsto (A_M,u^M_\Gamma)
\end{equation*}
is called \emph{momentum space Feynman rules}, and the map
\begin{equation*}
\Gamma\mapsto (A_P,u^P_\Gamma)
\end{equation*}
is called \emph{position space Feynman rules}.\\

Usually, in the physics literature, the restriction to the subspace is imposed by multiplying the direct product
of propagators with several delta distributions which are interpreted as ''momentum conservation'' at each vertex
in the momentum space picture, and dually ''translation invariance'' in the position space case.\\

In position space, it is immediately seen that
\begin{equation*}
u_\Gamma^P=(j_\Gamma^{\oplus d})^\ast u_{0,P}^{\otimes |E(\Gamma)|} = \pi_\ast \prod_{e\in E(\Gamma)} u_{0,P}(x_{e,out}-x_{e,in})
\end{equation*}
where $\pi_\ast$ means pushforward along the projection $\pi:M^{V(\Gamma)\vee}\rightarrow M^{V(\Gamma)\vee}/H^0(\Gamma)^d.$
\cite{BBK}. \\

In momentum space, things are a bit more complicated.
\begin{dfn} A connected graph $\Gamma$ is called \emph{core} if $\operatorname{rk} H_1(\Gamma\setminus \{e\})<\operatorname{rk} H_1(\Gamma)$ for all $e\in E(\Gamma).$
\end{dfn}
By Euler's formula (which follows from the exactness of
(\ref{eq:exactseq}))
\begin{equation*}
\rk H_1(\Gamma)-|E(\Gamma)|+|V(\Gamma)|-\rk H_0(\Gamma)=0,
\end{equation*}
it is equivalent for a connected graph $\Gamma$ to be core and to be one-particle-irreducible (1PI), a physicists' notion:
$\Gamma$ is one-particle-irreducible if removing an edge does not disconnect $\Gamma.$ \\

Let now $\Gamma$ be connected and
core, then
\begin{equation*}
u_\Gamma^M=(i_\Gamma^{\oplus d})^\ast u_{0,M}^{\otimes |E(\Gamma)|} = \prod_{e\in E(\Gamma)} u_{0,M}(p_e) \prod_{v\in V(\Gamma)} \delta_0(\sum_{e\in E(\Gamma)} (v:e)p_e).
\end{equation*}
This is simply because $\operatorname{im} i_\Gamma=\ker\partial,$ and because for
\begin{equation*}
\partial(\sum_{e\in E(\Gamma)} p_e e)=\sum_{e\in E(\Gamma)}p_e\sum_{v}(v:e)v=0
\end{equation*}
it is necessary that
\begin{equation*}
\sum_{e\in E(\Gamma)} (v:e)p_e=0
 \mbox{ for all }v\in V(\Gamma).
 \end{equation*}
(The requirement that $\Gamma$ be core is really needed here because otherwise certain $e\in E(\Gamma)$ would never show up
in a cycle, and hence would be missing inside the delta function.)\\

Moreover, one can define a version of $u_\Gamma^M$ which depends additionally on \emph{external momenta }$P_v\in\mathcal{M},$
one for each $v\in V(\Gamma),$ up to momentum conservation for each component $\sum_{v\in C} P_v=0:$
\begin{equation}\label{eq:ugammambig}
U_\Gamma^M(\{P_v\}_{v\in V(\Gamma)})=\prod_{e\in E(\Gamma)} u_{0,M}(p_e) \prod_{v\in V(\Gamma)} \delta_0(P_v+\sum_{e\in E(\Gamma)} (v:e)p_e).
\end{equation}
By a slight abuse of notation I keep the $P_v,$ $v\in V(\Gamma),$ as coordinate vectors for $\mathcal{M}^{|V(\Gamma)|}/H^0(\Gamma,\R)^d=A_P$ and identify distributions
on $A_P$ with distributions on $\mathcal{M}^{|V(\Gamma)|}$ that are multiples of $\prod_C \delta_0(\sum_{v\in C}P_v).$ \\

$U_\Gamma^M$ is now a distribution on a subset of $A_P\times A_M,$ and
\begin{equation*}
U_\Gamma^M|_{P_v=0,v\in V(\Gamma)}=u_\Gamma^M.
\end{equation*}
 The vectors in $P_v\in A_P$ determine a shift of
the linear subspace $A_M=H_1(\Gamma;\R)^{\oplus
d}\hookrightarrow \mathcal{M}^{|E(\Gamma)|}$ to an affine one.
Usually all but a few of the $P_v$ are set to zero, namely all but
those which correspond to the incoming or outgoing particles of an experiment
(see section \ref{ss:physics}). \\

The relation between the momentum space and position space distributions is then a Fourier duality. I denote by $\mathcal{F}$ the Fourier
transform.

\begin{pro}\label{pro:Fourier}
If the basic propagators
are Fourier-dual $(\mathcal{F}u_{0,P}=u_{0,M}),$ as is the case for $u_{0,M}(p)=\frac{1}{|p|^2}$ and $u_{0,P}(x)=\frac{1}{|x|^{d-2}},$ then
\begin{equation*}
(U_\Gamma^M[\underline{1}_{A_M}])(\{P_v\})=\mathcal{F} u_\Gamma^P
\end{equation*}
where only the (internal) momenta of $A_M$ are integrated out; and this holds
up to convergence issues only, i.~e.~ in the sense of \dref{dfn:integral}.
\end{pro} \hfill $\Box$

For example, the graph
\begin{equation*}
\Gamma_3=\gammafive
\end{equation*}
gives rise to
\begin{eqnarray*}
u^{M}_{\Gamma_3} & = & u_{0,M}^2(p_1)u_{0,M}(p_2)u_{0,M}(p_1+p_2)u_{0,M}(p_3)u_{0,M}(p_2+p_3),\\
u^{P}_{\Gamma_3} & = & u_{0,P}(x_1-x_2)u_{0,P}(x_1-x_3)u_{0,P}(x_2-x_3)u_{0,P}(x_2-x_4)u_{0,P}^2(x_3-x_4),
\end{eqnarray*}
where $p_1^i,\ldots,p_3^i,$ $i=0,\ldots,d-1$ is a basis of coordinates for $A_M$ and $x_1^i,\ldots,x_4^i,$ $i=0,\ldots,d-1$ is a basis of coordinates
for $\mathcal{M}^{V(\Gamma_3)\vee}$ (If $\Gamma$ is connected, dividing by $H^0(\Gamma;\R)^d$ takes care of the joint (diagonal) translations by $\mathcal{M}$ and, as previously,
instead of writing distributions on $\mathcal{M}^{V(\Gamma)\vee}/H^0(\Gamma;\R)^d,$ I take
the liberty of writing translation-invariant
distributions on $\mathcal{M}^{V(\Gamma)\vee}).$\\

Finally the case of external momenta:
\begin{eqnarray}\label{eq:p1p2p4}
U^{M}_{\Gamma_3}(P_1,P_2,0,P_4) &=&  u_{0,M}(p_1)u_{0,M}(p_1+P_1)u_{0,M}(p_2)u_{0,M}(p_1+p_2+P_1+P_2)\nonumber\\
&&\times u_{0,M}(p_3)u_{0,M}(p_2+p_3+P_4)\delta_0(P_1+P_2+P_4).
\end{eqnarray}

I set one of the external momenta, $P_3,$ to zero in order to
have a constant number of 4 adjacent (internal and external)
momenta at each vertex: $P_1$ is the sum of two external momenta at the vertex $1$ (See
section \ref{ss:physics} for the reason). \\

We will come back to the question of the affine subspaces $H_i$ where $u^{M}_\Gamma$ resp.~ $u^{P}_\Gamma$ is not defined
in the section about renormalization.\\

In general, following \cite[Section 2]{BEK}, a configuration is just an inclusion of a vector space $W$ into another vector
space $\R^E$ with fixed basis $E:$ The dual basis vectors $e^\vee,$ $e\in E$ determine linear forms on $W,$ and those
linear forms (or dually the linear hyperplanes annihilated by them) are the ''points'' of the configuration in the usual sense.
By the above construction, any such configuration, plus the choice of a propagator, defines an integral.

If the configuration comes from a Feynman graph, the integral is called \emph{Feynman integral}.
\subsection{Parametric representation}\label{ss:schwinger}
Integrals can be rewritten in many ways, using linearity of the integrand, of the domain, change of variables and Stokes' theorem,
and possibly a number of other tricks.

For many purposes it will be useful to have a version of the Feynman rules with a domain $A$ which is much lower-dimensional than
in the previous section but has boundaries and corners. The first part of the basic trick here is to rewrite
the propagator
\begin{equation*}
u_0 = \int_0^\infty \operatorname{exp}(-a_e u_0^{-1})da_e
\end{equation*}
(whenever the choice of propagator allows this inversion; $u_0(p)=\frac{1}{|p|^{2}}$ certainly does), introducing a
new coordinate $a_e\in \R_{\ge 0}$ for each edge $e\in E(\Gamma).$
Like this one has a distribution
\begin{equation}\label{eq:u0e}
\bigotimes_{e\in E(\Gamma)} \exp(-a_e u_0^{-1}(p_e))=\exp\left(-\sum_{e\in E(\Gamma)}a_e u_0^{-1}(p_e)\right)
\end{equation}
in $(\mathcal{M}\times \R_{\ge 0})^{|E(\Gamma)|}.$
From now on I assume $u_0(p)=\frac{1}{|p|^2}.$ Suppose $i:
W\hookrightarrow \R^{|E(\Gamma)|}$ is an inclusion. Once a
basis of $W$ is fixed, the linear form $e^\vee i$ is a row
vector in $W$ and its transpose $(e^\vee i)^t$ a column vector
in $W.$ The product $(e^\vee i)^t(e^\vee i)$ is then a $\dim
W$-square matrix. Pulling back (\ref{eq:u0e}) along an
inclusion $i^{\oplus d}:W\hookrightarrow
\mathcal{M}^{|E(\Gamma)|}$ (such as $i^{\oplus
d}=i_\Gamma^{\oplus d}$ or $i^{\oplus d}=j_\Gamma^{\oplus d})$
means imposing linear relations on the $p_e.$ These relations
can be transposed onto the $a_e:$ After integrating gaussian
integrals over $W$ (this is the second part of the trick) and a
change of variables, one is left with the distribution
\begin{equation*}
u_\Gamma^S(\{a_e\}) = \left( \det \sum_{e\in E(\Gamma)} a_e (e^\vee i)^t (e^\vee i) \right)^{-d/2}
\end{equation*}
on $A_S=\R^{|E(\Gamma)|}_{\ge 0}$ except certain intersections
$H_i$ of coordinate hyperplanes $\{a_e=0\}.$ I discarded a multiplicative
constant $C_\Gamma=(2\pi)^{d\dim W/2}$ which does not depend on the
topology of the graph.  \\

Suppose that $d=4.$ Depending on whether $i=i_\Gamma$ or
$j_\Gamma$ there is a momentum space and a position space
version of this trick. The two are dual to each other in the
following sense:
\begin{equation*}
\det \sum_{e\in E(\Gamma)} a_e (e^\vee i_\Gamma)^t (e^\vee i_\Gamma) = \left(\prod_{e\in E(\Gamma)} a_e\right)
\det \sum_{e\in E(\Gamma)} a_e^{-1} (e^\vee j_\Gamma)^t (e^\vee j_\Gamma)
\end{equation*}
See \cite[Proposition 1.6]{BEK} for a proof. In this paper, we
will only consider the momentum space version, where
$i=i_\Gamma.$ The map
\begin{equation*}
\Gamma\mapsto (A_S,u_\Gamma^S)
\end{equation*} with $i=i_\Gamma$ is called \emph{Schwinger} or \emph{parametric Feynman rules}.
Just as in the previous section, there is also a version with
external momenta which I just quote from
\cite{IZ,BlKr,BlochJapan}:
\begin{equation*}
U_\Gamma^S(\{a_e\},\{P_v\}) = \frac{\exp(-(N^{-1}P)^tP)}{\left(\det \sum_{e\in E(\Gamma)} a_e (e^\vee i_\Gamma)^t (e^\vee i_\Gamma) \right)^2}
\end{equation*}
where
\begin{equation*}
N = \sum_{e\in E(\Gamma)} a_e^{-1} (e^\vee j_\Gamma)^t (e^\vee j_\Gamma),
\end{equation*}
a $d(|V(\Gamma)|-\dim H_0(\Gamma;\R))$-square matrix.\\

The determinant
\begin{equation*}
\Psi_\Gamma(a_e) = \det \sum_{e\in E(\Gamma)} a_e (e^\vee i_\Gamma)^t (e^\vee i_\Gamma)
\end{equation*}
is a very special polynomial in the $a_e.$ It is called
\emph{first graph polynomial}, \emph{Kirchhoff polynomial} or
\emph{Symanzik polynomial}. It can be rewritten
\begin{equation}\label{eq:graphpoly}
\Psi_\Gamma(a_e) = \sum_{T \operatorname{sf} \operatorname{of} \Gamma} \prod_{e\not\in E(T)} a_e
\end{equation}
as a sum over \emph{spanning forests} $T$ of
$\Gamma:$ A spanning forest is a subgraph $E(T)\subseteq
E(\Gamma)$ such that the map $\partial|_{\R^{E(T)}}:
\R^{E(T)}\rightarrow \R^{V(\Gamma)}/H_0(\Gamma;\R)$ is an
isomorphism; in other words, a subgraph without cycles that has
exactly the same components as $\Gamma.$ (In the special case
where $\Gamma$ is connected, a spanning forest is called a
\emph{spanning tree} and is characterized by being connected as
well and having no
cycles.) \\

For the \emph{second graph polynomial} $\Phi_\Gamma,$ which is
a polynomial in the $a_e$ and a quadratic form in the $P_v,$
let us assume for simplicity that $\Gamma$ is connected. Then
\begin{equation*}
\Phi_\Gamma(a_e,P_v)=\Psi_\Gamma\cdot (N^{-1}P)^t P = \sum_{T \operatorname{st} \operatorname{of} \Gamma}\sum_{e_0\in E(T)} P_1^tP_2 a_{e_0}\prod_{e\not\in E(T)} a_e
\end{equation*}
where $P_A=\sum_{v\in C_A} P_v$ is the sum of momenta in the
first connected component $C_A$ and $P_B=\sum_{v\in C_B}
P_v$ the sum of momenta in the second connected component $C_B$ of the graph
$E(T)\setminus \{e_0\}$ (which has exactly two components since $T$ is a spanning tree).
See \cite{BEK,BlKr,IZ} for proofs.\\

Here is a simple example: If
\begin{equation*}
\Gamma_2=\gammaeiner
\end{equation*}
then
\begin{eqnarray*}
\Psi_{\Gamma_2}&=&a_1+a_2\\
\Phi_{\Gamma_2}&=&P_1^2a_1a_2
\end{eqnarray*}
and
\begin{equation*}
U_\Gamma^S = \frac{\exp\left(-P_1^2\frac{a_1a_2}{a_1+a_2}\right)}{(a_1+a_2)^2}.
\end{equation*}
All this holds if $u_{0,M}=\frac{1}{|p|^2}.$ If $u_{0,M}=\frac{1}{|p|^2+m^2}$ then
\begin{equation*}
U_\Gamma^S = \exp(-m^2\sum_{e\in E(\Gamma)}a_e)U_\Gamma^S|_{m=0}.
\end{equation*}
\subsection{The origin of Feynman graphs in physics}\label{ss:physics}
Before we continue with a closer analysis of the divergence
locus of these Feynman integrals, it will be useful to have at
least a basic understanding of why they were introduced in
physics. See
\cite{Weinberg,Collins,KBook,Fredenhagen,Haag,RS2,Veltman,Cvitanovic},
for a general exposition,
and I follow in particular \cite{RS2,Fredenhagen} in this
section. Quantum Field Theory is a theory of particles which
obey the basic principles of quantum mechanics and special
relativity at the same time. Special relativity is essentially
the study of the Poincar\'e group
\begin{equation*}
\mathcal{P} = \R^{1,3} \rtimes \operatorname{SL}(2,\C)
\end{equation*}
(where $\operatorname{SL}(2,\C)\rightarrow O(1,3)^+$ is the
universal double cover of the identity component $O(1,3)^+$ of
$O(1,3)).$ In other words, $\mathcal{P}$ is the double cover of
the group of (space- and time-) orientation-preserving
isometries of Minkowski space-time $\R^{1,3}$ (I assume $d=4$
in this section). \\

On the other hand, quantum mechanics always comes with a
Hilbert space, a vacuum vector, and operators on
the Hilbert space. \\

By definition, a \emph{single particle} is then an irreducible
unitary representation of $\mathcal{P}$ on some Hilbert space
$H_1.$ Those have been classified by Wigner according to the
joint spectrum of $P=(P_0,\ldots,P_3),$ the vector of
infinitesimal generators of the translations: Its joint
spectrum (as a subset of $\R^{1,3}$) is either one of the
following $\operatorname{SL}(2,\C)$-orbits: the hyperboloids
(mass shells) $S_\pm(m)=\{(p^0)^2-(p^1)^2-(p^2)^2-(p^3)^2=m^2,
p^0\gtrless 0\}\subset \R^{1,3},$ $(m>0),$ and the forward- and
backward lightcones $S_\pm(0)\subset \R^{1,3}$ $(m=0).$ (There
are two more degenerate cases, for example $m<0$ which I don't
consider further.) This gives a basic distinction between
massive $(m>0)$ and massless particles $(m=0).$ For a finer
classification, one looks at the stabilizer subgroups $G_p$ at
$p\in S_\pm(m).$ If $m>0,$ $G_p\cong SU(2,\C),$ If $m=0,$ $G_p$
is the double cover of the group of isometries of the euclidean
plane. In any case, the $G_p$ are pairwise conjugate in
$\operatorname{SL}(2,\C)$ and
\begin{equation*}
H_1 = \int_{\oplus} H^p d\Omega_m(p)
\end{equation*}
where the $H^p$ are pairwise isomorphic and carry an
irreducible representation of $G_p.$ By $d\Omega_m$ I denote
the unique $\operatorname{SL}(2,\C)$-invariant measure on
$S_\pm.$  The second classifying parameter
is then an invariant of the representation of $G_p$ on $H^p:$
In the case where $m>0$ and $G_p\cong SU(2,\C),$ one can take
the dimension: $H^p \cong \C^{2s+1},$ and $s\in \N/2$ is called
\emph{spin}. If $m=0,$ $G_p$ acts on $\C$ by mapping a rotation
by the angle $\phi$ around the origin to $e^{i n\phi}\in
\C^\ast,$ and $n/2$ is called \emph{helicity} (again I dismiss
a few cases which are
of no physical interest). \\

In summary, one identifies a single particle of mass $m$ and
spin $s$ or helicity $n$ with the Hilbert space
\begin{equation*}
H_1 \cong L_2(S_\pm(m),d\Omega_m)\otimes \C^{2s+1} \mbox{ resp. }
L_2(S_\pm(0),d\Omega_0),
\end{equation*}
and a \emph{state} of the given particle is an element of the
projectivized Hilbert space $\mathbb{P}H_1.$ \\

 Quantum field theories describe many-particle systems, and
particles can be generated and annihilated. A general result in
quantum field theory, the Spin-Statistics theorem [62, 55],
tells that systems of particles with integer spin obey Bose
(symmetric) statistics while those with half-integer spin obey
Fermi (antisymmetric) statistics. We stick to the case of
$s=0,$ and most of the time even $m=0,$ $n=0,$ (which can be
considered as a limit $m\rightarrow 0$ of the massive case) in
this paper.\\

The Hilbert space of infinitely many non-interacting particles
of the same type, called Fock space, is then
\begin{equation*}
H = \operatorname{Sym} H_1 = \bigoplus_{n=0}^\infty \operatorname{Sym}^n H_1
\end{equation*}
the symmetric tensor algebra of $H_1$ (For fermions, one would
use the antisymmetric tensor algebra). $\mathcal{P}$ acts on
$H$ in the obvious way, denote the representation by $U,$ and
$\Omega=1\in \C=\operatorname{Sym}^0 H_1\subset H$ is called
\emph{vacuum vector}. \\

Particles are created and annihilated as follows: If $f\in
\mathcal{D}(\R^{1,3})$ is a test function, then $\hat f =
\mathcal{F}f|_{S_\pm(m)} \in H_1,$ (the Fourier transform is
taken with respect to the Minkowski metric) and
\begin{eqnarray*}
a^\dagger[f]: \operatorname{Sym}^{n-1} H_1&\rightarrow &\operatorname{Sym}^{n}H_1:\\
 \Phi(p_1,\ldots,p_{n-1})&\mapsto& \sum_{i=1}^n
\hat f(p_i)\Phi(p_1,\ldots,\widehat{p_i},\ldots,p_n)\\
a[f]:\operatorname{Sym}^{n+1} H_1&\rightarrow &\operatorname{Sym}^{n}H_1:\\
\Phi(p_1,\ldots,p_{n+1})&\mapsto &\int_{S_\pm(m)}
\overline{\hat f(p)}\Phi(p,p_1,\ldots,p_n)d\Omega_m(p),\\
\end{eqnarray*}
define operator-on-$H$-valued distributions $f\mapsto
a^\dagger[f],$ $f\mapsto a[f]$ on $\R^{1,3}.$ The operator
$a^\dagger[f]$ creates a particle in the state $\hat f$ (i.~e.~
with smeared momentum $\hat f$), and $a[f]$ annihilates one.
\\

 The sum
\begin{equation*}
\phi=a+a^\dagger
\end{equation*}
 is called \emph{field.} It is the quantized
version of the classical field, which is a $C^\infty$ function
on Minkowski space. The field $\phi$ on the other hand is an
operator-valued distribution on Minkowski space. It satisfies
the Klein-Gordon equation
\begin{equation}\label{eq:kg}
(\Box+m^2)\phi=0
\end{equation}
$(\Box$ is the Laplacian of $\R^{1,3})$ which is the
Euler-Lagrange equation for the classical Lagrangian
\begin{equation}\label{eq:lagr}
\mathcal{L}_{0} = \frac{1}{2}(\partial_\mu \phi)^2-\frac{1}{2}m^2 \phi^2.
\end{equation}
The tuple $(H,U,\phi,\Omega)$ and one extra datum which I omit
here for simplicity is what is usually referred to as a quantum
field theory satisfying the Wightman-axioms
\cite{SW}. The axioms require certain $\mathcal{P}$-equivariance,
continuity and \emph{locality} conditions.\\

The tuple I have constructed (called the \emph{free} scalar
field theory) is a very well understood one because
(\ref{eq:kg}) resp.~ the Lagrangian (\ref{eq:lagr}) are very
simple indeed. As soon as one attempts to construct a quantum
field theory $(H_I,U_I,\phi_{I},\Omega_{I})$ for an interacting
Lagrangian (which looks more like a piece of the Lagrangian of
the Standard model) such as
\begin{equation}\label{eq:lag}
\mathcal{L}_{0}+\mathcal{L}_I = \frac{1}{2}(\partial_\mu \phi_I)^2-\frac{1}{2}m^2 \phi_I^2+\lambda \phi_I^n,
\end{equation}
$(n\ge 3,\lambda\in\R$ is called \emph{coupling constant}) one
runs into serious trouble. In this rigorous framework the
existence and construction of non-trivial interacting quantum
field theories in four dimensions is as of today an unsolved
problem, although there is an enormous number of important
partial results, see for example \cite{Rivasseau}.\\

However, one can expand quantities of the interacting quantum
field theory as a formal power series in $\lambda$ with
coefficients quantities of the free field theory, and hope that
the series has a positive radius of convergence. This is called
the \emph{perturbative expansion}. In general the power series
has radius of convergence 0, but due to some non-analytic
effects which I do not discuss further, the first terms in the
expansion do give a very good approximation to the
experimentally observed quantities for many important
interacting theories (this is the reason why quantum field
theories play such a prominent role in the physics of the last
50 years).\\

I will devote the remainder of this section to a sketch of this
perturbative expansion, and how the Feynman integrals
introduced in the previous section arise there.\\

By Wightman's reconstruction theorem \cite{SW}, a quantum
field theory $(H_I,U_I,\phi_I,\Omega_I)$ is uniquely determined
by and can be reconstructed from the \emph{Wightman functions}
(distributions) $w^I_n=\left< \Omega_I, \phi_I(x_1)\ldots
\phi_I(x_n)\Omega_I\right>.$ Similar quantities are the
\emph{time-ordered Wightman functions}
\begin{equation*}
t^I_n= \left<\Omega_I,T(\phi_I(x_1)\ldots \phi_I(x_n))\Omega_I\right>
\end{equation*}
which appear directly in scattering theory. If one knows all
the $t^I_n,$ one can compute all scattering cross-sections. The
symbol $T$ denotes time-ordering:
\begin{eqnarray*}
T( \psi_1(x_1)\psi_2(x_2)) &=& \psi_1(x_1)\psi_2(x_2) \mbox{ if }x_1^0 \ge x_2^0\\
&=& \psi_2(x_2)\psi_1(x_1) \mbox{ if }x_2^0> x_1^0
\end{eqnarray*}
for operator-valued distributions $\psi_1,\psi_2.$ \\

For the free field theory, all the $w_n$ and $t_n$ are
well-understood, in particular
\begin{eqnarray*}
t_2(x_1,x_2) &=& \left<\Omega,T(\phi(x_1)\phi(x_2))\Omega\right>\\
&=& \mathcal{F}^{-1}\frac{i}{(p^0)^2-(p^1)^2-(p^2)^2-(p^3)^2-m^2+i\epsilon}
\end{eqnarray*}
where the Fourier transform is taken with respect to the
difference coordinates $x_1-x_2$ (the $t_n$ are
translation-invariant). $t_2$ is a particular fundamental
solution of equation (\ref{eq:kg}) called the
\emph{propagator}. By a technique called \emph{Wick rotation},
one can go forth and back between Minkowski space $\R^{1,3}$
and euclidean $\R^4 $ \cite{OS,Kazhdan}, turning Lorentz
squares $(p^0)^2-(p^1)^2-(p^2)^2-(p^3)^2$ into euclidean
squares $-|p|^2,$ and the Minkowski space propagator $t_2$ into
the distribution $u_{0,P}=\mathcal{F}^{-1}\frac{1}{|p|^2+m^2}$ introduced in
the previous sections. In the massless case $m=0,$ we have
$u_{0,P}=u_{0,M}=\frac{1}{|x|^2}$ if $d=4.$\\

From the usual physics axioms for scattering theory and on a
purely symbolic level, Gell-Mann's and Low's formula relates
the interacting $t^I_n$ with vacuum expectation values
$\left<\Omega,T(\ldots)\Omega\right>$ of time-ordered products
of powers of the \emph{free} fields
\begin{equation}\label{eq:gml}
t^I_n(x_1,\ldots,x_n) = \sum_{k=0}^\infty\frac{i^k}{k!} \int \left< \Omega,T(\phi(x_1)\ldots\phi(x_n)\mathcal{L}_I^0(y_1)\ldots
\mathcal{L}_I^0(y_k))\Omega\right> d^4y_1\ldots d^4y_k
\end{equation}
as a formal power series in $\lambda.$ I denote
$\mathcal{L}_I^0 = \mathcal{L}_I|_{\phi_I\rightarrow
\phi}=\lambda\phi^n.$ (There is a subtle point here in defining
powers of $\phi$ as operator-valued distributions. The solution
is called \emph{Wick powers}: In $\phi^n=(a+a^\dagger)^n,$ all
monomials containing $aa^\dagger$ in this order are discarded.)
But now within the free field theory, the $\left<
\Omega,T(\ldots)\Omega\right>$ are well-understood: It follows
from the definition of $T,a,a^\dagger$ and the Wick powers that
$\left< \Omega,T(\ldots)\Omega\right>$ is a polynomial in the
$t_2,$ more precisely
\begin{equation}\label{eq:top}
\left< \Omega,T(\phi^{n_1}\ldots\phi^{n_k})\Omega\right>= \sum_\Gamma c_\Gamma \pi^\ast u_\Gamma^P.
\end{equation}
where the sum is over all Feynman graphs $\Gamma$ with $k$
vertices such that the $i$th vertex has degree $n_i,$ and where
$u_\Gamma^P$ is defined as in the previous sections, $c_\Gamma$
a combinatorial symmetry factor, and $u_{0,P}(x)=t_2(x,0)$ up to a Wick rotation. \\

If one uses (\ref{eq:top}) for (\ref{eq:gml}) then one gets
Feynman graphs with $n$ external vertices of degree 1. The
external edges, i.~e.~ edges leading to those $n$ vertices,
appear simply as tensor factors, and can be omitted (amputated)
in a first discussion. Like this
we are left with the graphs considered in the previous section.\\

It follows in particular that only Feynman graphs with vertices
of degree  $n$ appear from the
Lagrangian (\ref{eq:lag}). Note that whereas external
physical particles are always on-shell (i.~e.~ their momentum
supported on $S_\pm),$ the internal virtual particles are
integrated over all of momentum space in the Gell-Mann-Low formula. \\

In summary, the perturbative expansion of an interacting
quantum field theory (whose existence let alone construction in
the sense of the Wightman axioms is an unsolved problem)
provides an power series approximation in the coupling constant
to the bona fide interacting functions $t^I_n.$ The
coefficients are sums of Feynman integrals which are composed
of elements of the free theory only.

\section{Regularization and renormalization}\label{s:regren}
The Feynman integrals introduced so far are generally divergent integrals. At first sight it
seems to be a disturbing feature of a quantum field theory that it produces divergent integrals
in the course of calculations, but a closer look reveals that this impression is wrong: it is
only a naive misinterpretation of perturbation theory that makes us think that way.\\

Key to this is the insight that single Feynman graphs are really about virtual particles,
and their parameters, for example their masses, have no real physical meaning. They have to
be \emph{renormalized}. Like this the divergences are compensated by so-called counterterms in the
Lagrangian of the theory which provide some kind of dynamical contribution to those parameters
\cite{Collins}. I will not make further use of this physical interpretation but only consider
mathematical aspects. If the divergences can be compensated by adjusting only a finite number of
parameters in the Lagrangian (i.~e.~ by leaving the form of the Lagrangian invariant and
not adding an infinity of new terms to it) the theory is called renormalizable. \\

An important
and somehow nontrivial, but fortunately solved \cite{BS,BPH,Zimmermann,EG,Kreimer,CK,CK2}, problem is to find a way to organize this correspondence
between removing divergences and compensating counterterms in the Lagrangian for arbitrarily complicated graphs.
Since the terms in the Lagrangian are local terms, that is polynomials in the field and its
derivatives, a necessary criterion for this is the so-called \emph{locality of counterterms}: If
one has a way of removing divergences such that the correction terms are local ones, then
this is a good indication that they fit into the Lagrangian in the first place.\\

\emph{Regularization} on the other hand is the physics term
used for a variety of methods of writing the divergent integral
or integrand as the limit of a holomorphic family of convergent
integrals or integrands, say over a punctured disk. Sometimes
also the integrand is fixed, and the domain of integration
varies holomorphically say over the punctured disk. We will see
a number of such regularizations in the remainder of this
paper.
\subsection{Position space}\label{ss:positionspace}
In position space, the renormalization problem has been known for a long time to be an \emph{extension problem of distributions}
\cite{BS,EG}. This follows already from our description in
section \ref{s:feynman}, but it will be useful to have a closer look at the problem.
Recall the position space Feynman distribution
\begin{equation*}
u^P_{\Gamma}=(j^{\oplus d}_\Gamma)^\ast u_{0,P}^{\otimes |E(\Gamma)|}
\end{equation*}
is defined only as a
distribution on
$A_P=\mathcal{M}^{|V(\Gamma)|\vee}/H^0(\Gamma;\R)^{\oplus d}$
minus certain affine (in this case even linear) subspaces.
Suppose for example
\begin{equation*}
\Gamma_2=\gammaeiner
\end{equation*}
with $u_{\Gamma_2}^P= \frac{1}{|x|^{2d-4}}.$ If $f$ is a
non-negative test function supported in a ball $N=\{|x|\le
\epsilon\}$ around 0.
\begin{equation*}
u^P_{\Gamma_2}[f] = \int_N f(x)u^P_{\Gamma_2}(x) dx \ge \operatorname{min}_{x\in N}f(x)\int d\Omega \int_0^\epsilon \frac{dr^{d-1}}{r^{2d-4}}.
\end{equation*}
If $d-1-(2d-4)\le -1,$ that is $d\ge 4,$ the integral will be
divergent at $0$ and $u^P_{\Gamma_2}$ not defined on test
functions supported at $0.$ This is the very nature of
ultraviolet (i.~e.~ short-distance) divergences. On the other
hand, divergences as some position-space coordinates go to
$\infty,$ are called infrared (long-distance) divergences. We
will be concerned with ultraviolet divergences in this paper. \\

For simplicity we restrict ourselves to graphs with at most logarithmic divergences throughout the rest of the paper, that is
$d \rk H_1(\gamma)\ge 2|E(\gamma)|$ for all subgraphs $E(\gamma)\subseteq E(\Gamma).$ A subgraph $\gamma$ where equality holds is called \emph{divergent}.
A detailed power-counting analysis, carried out in \cite{BBK} shows that $u^P_{\Gamma}$ is only defined as a distribution inside
\begin{equation}\label{eq:divarr}
A_P^\circ = A_P \setminus \bigcup_{\onatop{E(\gamma)\subseteq E(\Gamma)}{d \rk H_1(\gamma)=2|E(\gamma)|}} \bigcap_{e\in E(\gamma)} \pi D_e
\end{equation}
where $D_e=\{x_{e,out}-x_{e,in}=0\}.$ The singular support (the locus where $u^P_{\Gamma}$ is not smooth) is
\begin{equation*}
\operatorname{sing} \operatorname{supp}u^P_{\Gamma} = A_P^\circ\cap \bigcup_{e\in E(\Gamma)} \pi D_e .
\end{equation*}
An extension of $u_\Gamma^P$ from $A_P^\circ$ to $A_P$ is
called a \emph{renormalization} provided it satisfies certain
consistency conditions to be discussed later. \\

In the traditional literature, which dates back to a central
paper of Epstein and Glaser \cite{EG}, an extension of
$u^P_\Gamma$ from $A_P^\circ$ to all of $A_P$ was obtained
inductively, by starting with the case of two vertices, and
embedding the solution (extension) for this case into the
three, four, etc.~ vertex case using a partition of unity. Like
this, in each step only one extension onto a single point, say
0, is necessary, a well-understood problem with a
finite-dimensional degree of freedom: Two extensions differ by
a distribution supported at this point $0,$ and the difference
is therefore, by an elementary consideration, of the form
$\sum_{|\alpha|\le n} c_\alpha \partial^\alpha \delta_0$ with
$c_\alpha\in \C.$ Some of these parameters $c_\alpha$ are fixed
by physical requirements such as probability conservation,
Lorentz and gauge invariance, and more generally the
requirement that certain differential equations be satisfied by
the extended distributions. But even after these constants are
fixed, there are degrees of freedom left, and various groups
act on the space of possible extensions, which are collectively
called \emph{renormalization group.} For the at most logarithmic
graphs considered in this paper, $n=0$ and only one constant $c_0$ needs to
be fixed in each step.

\subsection{Resolution of singularities}\label{ss:ressing}

The singularities, divergences and extensions (renormalizations) of the Feynman distribution $u_\Gamma^P$
are best understood using a resolution of singularities \cite{BBK}. The Fulton-MacPherson compactification \cite{FM}
introduced in a quantum field theory context by Kontsevich \cite{KontsevichEMS,KontsevichTalk} and Axelrod and Singer \cite{AS}
serves as a universal smooth model where all position space Feynman distributions can be renormalized. In
\cite{BBK}, a graph-specific De Concini-Procesi Wonderful model \cite{deCPro} was used, in order to elaborate the
striking match between De Concini's and Procesi's notions of building set, nested set and notions found
in Quantum Field Theory. No matter which smooth model is chosen, one disposes of a smooth manifold $Y$ and
a proper surjective map, in fact a composition of blowups,
\begin{equation*}
\beta: Y\rightarrow A_P
\end{equation*}
which is a diffeomorphism on $\beta^{-1}(A_P^\circ)$ but $\beta^{-1}(A_P\setminus A_P^\circ)$ is
(the real locus of) a divisor with normal crossings. \\

Instead of the nonorientable smooth manifold $Y$ one can also
find an orientable manifold with corners $Y'$ and $\beta$ a
composition of real spherical blowups as in \cite{AS}. In my
pictures, the blowups are spherical because they are easier to
draw, but in the text they are projective. \\

Here is an example: If $\Gamma_3$ is again the graph
\begin{equation}\label{eq:examplegraph}
\Gamma_3 = \gammafive
\end{equation}
and $d=4$ then by (\ref{eq:divarr}) the locus where there are nonintegrable singularities is
\begin{equation*}
D_{1234} \subset D_{234} \subset D_{34}
\end{equation*}
where $D_{1234}=D_{12}\cap D_{13}\cap D_{14},$ $D_{234}=D_{23}\cap D_{24}.$ In $A_P,$ $\pi D_{1234}$ is
a point, $\pi D_{234}$ is 4-dimensional and $\pi D_{24}$ is 8-dimensional. Blowing up something means
replacing it by its projectivized normal bundle. The map $\beta$ is composed of three maps
\begin{equation*}
Y=Y_{34} \stackrel{\beta_3}{\rightarrow} Y_{234}\stackrel{\beta_2}{\rightarrow} Y_{1234}\stackrel{\beta_1}{\rightarrow} A_P
\end{equation*}
where $\beta_1$ blows up $D_{1234},$ $\beta_2$ blows up the strict transform of $D_{234},$ and $\beta_3$ blows up
the strict transform of $D_{34}.$
\begin{equation*}
\blowup \stackrel{\beta}{\rightarrow} \nest
\end{equation*}
Now $u_{\Gamma_3}^P$ can be pulled back along $\beta$ (because of lack of orientability of $Y,$ it will become
a distribution density). In a clever choice of local coordinates, for example
\begin{eqnarray*}
y_1^0&=& x_1^0-x_2^0\\
y_2^0&=& (x_2^0-x_3^0)/(x_1^0-x_2^0)\\
y_3^0&=& (x_3^0-x_4^0)/(x_2^0-x_3^0)\\
y_1^i&=& (x_1^i-x_2^i)/(x_1^0-x_2^0)\\
y_2^i&=& (x_2^i-x_3^i)/(x_2^0-x_3^0)\\
y_3^i&=& (x_3^i-x_4^i)/(x_3^0-x_4^0)
\end{eqnarray*}
one has
\begin{equation}\label{eq:wgammaP}
w_{\Gamma_3}^P = \beta^\ast u_{\Gamma_3}^P = \frac{f^P_{\Gamma_3}}{|y_1^0y_2^0y_3^0|}
\end{equation}
where $f_{\Gamma_3}^P$ is a locally integrable density which is even $C^\infty$ in the coordinates $y_1^0,y_2^0,y_3^0.$ The divergence is therefore isolated in the denominator,
and only in three directions: $y_1^0,$ $y_2^0$ and $y_3^0.$ The first is the local coordinate transversal to the exceptional divisor $\mathcal{E}_{1234}$
of the blowup of $D_{1234},$ the second transversal to the exceptional divisor $\mathcal{E}_{234}$ of the
blowup of $D_{234},$ and the third transversal to the exceptional divisor $\mathcal{E}_{34}$ of the blowup of $D_{34}$
(the difference between $\mathcal{E}_{34}$ and $D_{34}$ is not seen in the picture because of dimensional reasons). \\

For a general graph $\Gamma,$ the total exceptional divisor $\mathcal{E}=\beta^{-1}(A_P\setminus A_P^\circ)$ has normal
crossings and the irreducible components $\mathcal{E}_\gamma$ are indexed by connected divergent (consequently core) irreducible subgraphs $\gamma.$
Moreover,
\begin{equation*}
\mathcal{E}_{\gamma_1}\cap\ldots\cap\mathcal{E}_{\gamma_k} \neq \emptyset \Longleftrightarrow \mbox{ the }\gamma_i \mbox{ are nested}
\end{equation*}
where nested means each pair is either disjoint or one contained in the other. See \cite{BBK} for the
general result and more details.\\

Inspired by old papers of Atiyah \cite{Atiyah}, Bernstein and
Gelfand \cite{BG} we used $(u_\Gamma^P)^s,$ where $s$ in a
complex number in a punctured neighborhood of 1, as a
regularization \cite{BBK}. Similarly, since the propagator
$u_{0,P}(x)=\frac{1}{|x|^{d-2}}$ depends on the dimension, one
can also consider $u_\Gamma^P$ with $d$ in a punctured complex
neighborhood of 4 as a regularization but I will not pursue
this here.

\begin{dfn} A connected graph $\Gamma$ is called primitive if
\begin{equation*}
d \rk H_1(\gamma)= 2|E(\gamma)| \Longleftrightarrow E(\gamma)=E(\Gamma).
\end{equation*}
for all subgraphs $E(\gamma)\subseteq E(\Gamma).$
\end{dfn}
For a primitive graph $\Gamma_{p},$ only the single point $0\in A_P$ needs to be blown up, and the pullback along $\beta$ yields in suitable local coordinates
($y_1^0=x_1^0-x_2^0,$ $y_i^j=(x_i^j-x_{i+1}^j)/(x_1^0-x_2^0)$ otherwise)
\begin{equation*}
\beta^\ast u_{\Gamma_p}^P = \frac{f_{\Gamma_p}}{|y_1^0|}
\end{equation*}
where $f_{\Gamma_p}$ is a locally integrable distribution density constant in $y_1^0$-direction.
Let $d_\Gamma=d(|V(\Gamma_p)|-1).$
Consequently
\begin{equation*}
\beta^\ast (u_{\Gamma_p}^P)^s = \frac{f_{\Gamma_p}^s }{|y_1^0|^{d_{\Gamma_p} s-(d_{\Gamma_p}-1)}}
\end{equation*}
It is well-known that the distribution-valued function
$\frac{1}{|x|^s}$ can be analytically in a punctured neighborhood of
$s=1,$ with a simple pole at $s=1.$ The residue of this pole is
$\delta_0:$
\begin{equation*}
\frac{1}{|x|^s} = \frac{\delta_0}{s-1}+ |x|^s_{fin}, \quad |x|^s_{fin}[f]=\int_{-1}^1 |x|^s(f(x)-f(0))dx+\int_{\R\setminus [-1,1]}|x|^sf(x)dx.
\end{equation*}
This implies that the residue at $s=1$ of $\beta^\ast
(u_{\Gamma_p}^P)^s$ is a density supported
at the exceptional divisor (which is given in these coordinates
by $y_0=0,$ and integrating this density against the constant
function $\underline{1}_{Y}$ gives what is called the residue of the graph
$\Gamma_p$
\begin{equation*}
\operatorname{res}_P \Gamma_p = \operatorname{res}_{s=1} \beta^\ast (u_{\Gamma_p}^P)^s[\underline{1}_Y]= -\frac{2}{d_{\Gamma_p}}\int_{\mathcal{E}} f_{\Gamma_p}
\end{equation*}
(The exceptional divisor can actually be oriented in such a way
that $f_{\Gamma_p}$ is a degree
$(d_{\Gamma_p}-1)$ differential form).

Let us now come back to the case of $\Gamma_3$ which is not primitive but has a nested set of three divergent subgraphs.
Raising (\ref{eq:wgammaP}) to a power $s$ results in a pole at $s=1$ of order 3. The Laurent coefficient $a_{-3}$ of $(s-1)^{-3}$
is supported on
\begin{equation*}
\mathcal{E}_{1234}\cap\mathcal{E}_{234}\cap\mathcal{E}_{34},
\end{equation*}
for this is the set given in local coordinates by $y_1^0=y_2^0=y_3^0=0.$
Similarly, the coefficient of $(s-1)^{-2}$ is supported on
\begin{equation*}
(\mathcal{E}_{1234}\cap\mathcal{E}_{234})\cup(\mathcal{E}_{1234}\cap\mathcal{E}_{34})\cup(\mathcal{E}_{234}\cap\mathcal{E}_{34})
\end{equation*}
and the coefficient of $(s-1)^{-1}$ on
\begin{equation*}
\mathcal{E}_{1234}\cup\mathcal{E}_{234}\cup\mathcal{E}_{34}.
\end{equation*}
(The non-negative part of the Laurent series is supported everywhere on $Y$). Write $|dy|=|dy_1^0\ldots dy_3^3|.$ In order to compute the coefficient $a_{-3},$
one needs to integrate $f_{\Gamma_3},$ restricted to the subspace $y^0_1=y^0_2=y^0_3=0:$
\begin{eqnarray*}
f_{\Gamma_3}&=& \frac{|dy|}{(1+\underline{y}_1^2)(1+\underline{y}_2^2)(1+\underline{y}_3^2)}\\
&&\times \frac{1}{((1+y_2^0)^2+(\underline{y}_1+y_2^0\underline{y}_2)^2)((1+y_3^0)^2+(\underline{y}_2+y_3^0\underline{y}_3)^2)^)}
\end{eqnarray*}
where $\underline{y}_i$ denotes the 3-vector $(y_i^1,y_i^2,y_i^3).$ Consequently
\begin{equation*}
f_{\Gamma_3}|_{y_1^0=y_2^0=y_3^0=0}= \frac{|dy|}{(1+\underline{y}_1^2)^2
(1+\underline{y}_2^2)^2(1+\underline{y}_3^2)^2}=f_{\Gamma_1}^{\otimes 3}
\end{equation*}
 where $\Gamma_1$ is the primitive graph with two vertices and two parallel edges
joining them:
\begin{equation}\label{eq:gamma1}
\Gamma_1 = \gammaeiner
\end{equation}
The chart where (\ref{eq:wgammaP}) holds covers actually everything of $Y_P$ up to a set of measure
zero where there are no additional divergences. It suffices therefore to integrate in these coordinates only.
Several charts must be taken into account however when there are more than one maximal nested set.
In conclusion,
\begin{equation}\label{eq:stf}
a_{-3}[\underline{1}_Y]=(\operatorname{res}_P \Gamma_1)^3,
\end{equation}
a special case of a theorem in \cite{BBK}
relating pole coefficients of $\beta^\ast (u_\Gamma^P)^s$ to residues of graphs obtained from $\Gamma$ by contraction
of divergent subgraphs.\\

But the ultimate reason to introduce the resolution of
singularities in the first place is: In order to obtain an
extension (renormalization) of $u_\Gamma^P,$ one can now simply
remove the simple pole at $s=1$ along each component of the
exceptional divisor:
\begin{eqnarray}
w_{\Gamma_3}^P  &=& \frac{f_{\Gamma_3}}{|y_1^0y_2^0y_3^0|},\\
(w_{\Gamma_3}^P)_R &=& \frac{f_{\Gamma_3}}{|y_1^0|_{fin}|y_2^0|_{fin}|y_3^0|_{fin}}.
\end{eqnarray}
The second distribution $(w_{\Gamma_3}^P)_{R}$ is defined on
all of $Y,$ and consequently $\beta_\ast (w_{\Gamma_3}^P)_{R}$
on all of $A_P.$ It agrees with $u_{\Gamma_3}^P$ on test
functions having support in $A_P^\circ$ and is therefore an
extension. The difference between $w_{\Gamma_3}^P$ and
$(w_{\Gamma_3}^P)_R$ is a distribution supported on the
exceptional divisor which gives rise to a candidate for a counterterm
in the Lagrangian. \\

I call this renormalization scheme \emph{local minimal subtraction},
because locally, along each component of the exceptional divisor,
the simple pole is removed in a ''minimal way'', changing
only the principal part of the Laurent series. See
\cite{BBK} for a proof that this results in local counterterms,
a necessary condition for the extension to be a physically
consistent one.
\subsection{Momentum space}\label{ss:momentumspace}

In momentum space, the bad definition of the position space
Feynman distribution at certain diagonals $\bigcap D_e$ is
translated by a Fourier transform into ill-defined (divergent)
integrals with divergences at certain strata at infinity. For
example, the position space integral
$(\mathcal{M},u_{\Gamma_1}^P = u_{0,P}^2)$ in $d=4$ dimensions
for the graph
$\Gamma_1$ (see (\ref{eq:gamma1})) has a divergence at $0$ (which is the image $\pi D_{12}$ of the diagonal).
A formal Fourier transform would turn the pointwise product $u_{0,P}^2$ into a convolution product
\begin{equation*}
(\mathcal{F}u_{0,P}^2)(P)=\int u_{0,M}(p)u_{0,M}(p-P) d^4p.
\end{equation*}
In fact the right hand side is exactly $U_{\Gamma_1}^M(P)[\underline{1}_{A_{\Gamma_1}}]$ in agreement with \pref{pro:Fourier}. It does not converge at $\infty.$
(In order to see this we actually only need $U_{\Gamma_1}^M|_{P=0}=u_{\Gamma_1}^M,$ not the dependence upon external momenta). \\

On the other hand, the infrared singularities are to be found
at affine subspaces in momentum space. Of course the program
sketched in the previous section can be applied to the momentum
space Feynman distribution as well: A resolution of
singularities for the relevant strata at infinity can be found,
and the pullback of the momentum space Feynman distribution can
be extended onto all the irreducible components of the
exceptional divisor. But I want to use this section in order to
sketch another, algebraic,
approach to the momentum space renormalization problem, which is due to Connes and Kreimer \cite{Kreimer,CK,CK2}.\\

Assume $U_\Gamma^M[\underline{1}_{A_M}]$ varies holomorphically
with $d$ in a punctured disk around $d=4.$ Physicists call this
dimensional regularization \cite{Etingof,CoMaBook}: any integral $\int d^4p u(p)dp$ is
replaced by a $d$-dimensional integral $\int d^dp u(p)dp.$ Like
this we can consider $U_\Gamma^M$ as a distribution on \emph{all of}
$A_P\times A_M$ with values in
$\mathcal{R}=\C[[(d-4)^{-1},(d-4)]],$
the field of Laurent series in $d-4.$ If $U_\Gamma^M[f]$ is not convergent
in $d=4$ dimensions, then there will be a pole at $d=4.$\\

Let now $\sigma_\Gamma\in\mathcal{D}'(A_P)$ be a distribution with compact support. Since the distribution $U_\Gamma^M$ is smooth in the $P_v,$ we
can actually integrate it against the distribution $\sigma_\Gamma$ (For example, if $\sigma_\Gamma=\delta_0(|P_{v_1}|^2-E_1)\otimes\ldots\otimes\delta_0(|P^2_{v_n}|-E_n)$
then this amounts simply to evaluating $U_\Gamma^M$ at the subspaces $|P_{v_1}|^2=E_1,\ldots,|P_{v_2}|^2=E_n$). In any case we have a map
\begin{equation*}
\phi: (\Gamma,\sigma_\Gamma) \mapsto U_\Gamma^M[\underline{1}_{A_M}\otimes \sigma_\Gamma] \in \mathcal{R}
\end{equation*}
sending pairs to Laurent series.
Let now $\mathcal{H}$ be the polynomial algebra over $\C$ generated by isomorphism classes of connected core divergent graphs $\Gamma$
of a given renormalizable quantum field theory.
Define a coproduct $\Delta$ by
\begin{equation*}
\Delta(\Gamma)=1\otimes\Gamma+\Gamma\otimes1 + \sum_{\onatop{\gamma_1\sqcup\ldots\sqcup\gamma_k\subsetneq \Gamma}{\operatorname{conn.} \operatorname{core} \operatorname{div.} }} \gamma_1\cdots\gamma_k\otimes \Gamma//(\gamma_1\sqcup\ldots\sqcup\gamma_k).
\end{equation*}
The notation $\Gamma//\gamma$ means that any connected component of $\gamma$ inside $\Gamma$ is contracted to a (separate) vertex. By
standard constructions \cite{CK}, $\mathcal{H}$ becomes a Hopf algebra, called Connes-Kreimer Hopf algebra. Denote the antipode by $S.$
Let now $\mathcal{H}_\sigma$ be the corresponding Hopf algebra of pairs $(\Gamma,\sigma_\Gamma)$ (In order to define this Hopf algebra of
pairs, one needs the extra condition that $\sigma_\Gamma$ vanishes on all vertices that have no external edges, a standard assumption
if one considers only graphs of a fixed renormalizable theory). \\

The map $\phi: \mathcal{H}_\sigma\rightarrow\mathcal{R}$ is a
homomorphism of unital $\C$-algebras. The space of these maps $\mathcal{H}_\sigma\rightarrow \mathcal{R}$ is a group
with the convolution product $\phi_1\star\phi_2 = m(\phi_1\otimes\phi_2)\Delta.$ On $\mathcal{R}, $ there is the
linear projection
\begin{equation}\label{eq:r}
R: (d-4)^n \mapsto
\left\{\begin{array}{ll}0 & \mbox{ if } n\ge 0\\
(d-4)^n & \mbox{ if }n< 0\end{array}\right.
\end{equation}
onto the principal part.
\begin{thm}[Connes, Kreimer] The renormalized Feynman integral
$\phi_R(\Gamma,\sigma_\Gamma)|_{d=4}$ and the counterterm $S_R^\phi(\Gamma,\sigma_\Gamma)$ are given as follows. I denote $\underline{\Gamma}$ for the pair $(\Gamma,\sigma_\Gamma):$
\begin{eqnarray*}
S_R^\phi(\underline{\Gamma}) &=& -R\left(\phi(\underline{\Gamma})+\sum_{\onatop{\gamma=\gamma_1\sqcup\ldots\sqcup\gamma_k\subsetneq \Gamma}{\operatorname{conn.} \operatorname{core} \operatorname{div.} }} S_R^\phi(\underline{\gamma})\phi(\underline{\Gamma}//\underline{\gamma})\right)\\
\phi_R(\underline{\Gamma}) &=& (1-R)\left(\phi(\underline{\Gamma})+\sum_{\onatop{\gamma=\gamma_1\sqcup\ldots\sqcup\gamma_k\subsetneq \Gamma}{\operatorname{conn.} \operatorname{core} \operatorname{div.} }} S_R^\phi(\underline{\gamma})\phi(\underline{\Gamma}//\underline{\gamma})\right)
\end{eqnarray*} \hfill $\Box$
\end{thm}
These expressions are assembled from the formula for the antipode and the convolution product. Combinatorially, the Hopf algebra
encodes the BPHZ recursion \cite{BPH} and Zimmermann's forest formula \cite{Zimmermann}.
The theorem can be interpreted as a Birkhoff decomposition of the character $\phi$ into $\phi_-=S_R^\phi$ and $\phi_+=\phi_R$ \cite{CK2}. \\

The renormalization scheme described here is what I call \emph{global minimal subtraction}, because in
the target field $\mathcal{R},$ when all local information has been integrated out, the map $1-R$ removes only the
entire principal part at $d=4.$ This coincides with the renormalization scheme
described in \cite{Collins}. \\

In the case of $m=0$ and zero-momentum transfer (all but two external momenta set to 0) one knows that at $d=4$
\begin{equation}\label{eq:phir}
\phi_R(\Gamma) = \sum_{n=0}^N p_n(\Gamma) (\log |P|^2/\mu^2)^n, \quad p_n(\Gamma)\in \R
\end{equation}
where $\mu$ is an energy scale, and the $\sigma_\Gamma$ can be dropped for convenience. Let us now do our standard example
\begin{equation*}
\Gamma_3 = \gammafive
\end{equation*}
using the Hopf algebra. We interpret $\Gamma_3$ as a graph in $\phi^4$ theory,
so we think of two external edges at the first vertex, one at the second, and one at the fourth.
Recall the momentum space Feynman rules (\ref{eq:p1p2p4}) for $\Gamma_3.$ Let $P_2=0$ and write $P=P_1=-P_4$ such that
$P_1$ is the sum of the two external momenta entering at the first vertex.
Then
\begin{equation*}
\phi(\Gamma_3) = \int \frac{d^dp_1d^dp_2d^dp_3}{p_1^2(p_1+P)^2p_2^2(p_1+p_2+P)^2p_3^2(p_2+p_3-P)^2} \in \mathcal{R}.
\end{equation*}
This integral can be evaluated as a Laurent series in $d=4$ using standard techniques \cite{Collins}. It has a pole of order $3$ at
$d=4,$ and one might think of simply taking $(1-R)\phi(\Gamma_3)$ as a renormalized value, for this kills the principal
part, and the limit at $d=4$ may be taken. But the resulting counterterms would not be local ones, and the renormalization
would be physically inconsistent. The benefit of the Hopf algebra approach is that the necessary correction terms are
provided right away: \\

Let again $\gamma_1$ be the full subgraph with vertices 3 and 4, and $\gamma_2$ the full subgraph with vertices 2,3 and 4.
Then
\begin{eqnarray*}
\phi_R(\Gamma_3) &=& (1-R)\left.(\phi(\Gamma_3)-(R\phi(\gamma_2))\phi(\Gamma_3//\gamma_2)+\right.\\
&&+\left.R((R\phi(\gamma_1))\phi(\gamma_2//\gamma_1))\phi(\Gamma_3//\gamma_2)\right).
\end{eqnarray*}
Observe that, as a coincidental property of our example, $\Gamma_3//\gamma_2\cong\gamma_2//\gamma_1\cong\gamma_1$
(compare this with (\ref{eq:stf}),(\ref{eq:stf2})).\\

The Hopf algebra approach to renormalization has brought up a number of surprising connections
to other fields, see for example \cite{CK2,CK3,EFGK,MK2,FoissyBSM1,Sz1,Sz2,Sz3,Kreimergravity}. Other developments
starting from the Connes-Kreimer theory can be found in \cite{CoMaBook}.
Kreimer and van Suijlekom have shown that gauge and other symmetries
are compatible with the Hopf algebra structure \cite{Anatomy,Suijlekom1,Suijlekom2,vsmult,KreimerSui}. \\

A sketch how the combinatorics of the Hopf algebra relate to the resolution of singularities in the
previous section and to position space renormalization can be found in \cite{BBK}, see also section \ref{ss:misc}.

\subsection{Parametric representation}\label{ss:parametric}
In the parametric representation introduced in section \ref{ss:schwinger}, the divergences can be
found at certain intersections of the coordinate hyperplanes $A_e=\{a_e=0\}.$ This is in fact
one of the very reasons why the parametric representation was introduced: Consider for example the
divergent integral $(\R^4,u_{0,M}^2),$ with $u_{0,M}=\frac{1}{|p|^2},$
\begin{equation*}
\int \frac{d^4 p}{|p|^4} = \int \int_0^\infty \int_0^\infty \exp(-a_1|p|^2-a_2|p|^2) da_1 da_2 d^4p
\end{equation*}
in the sense of \dref{dfn:integral} (In this section, instead of $(A,u)$ I will simply write $\int_A u(x)dx.$)
The integral at the left hand side is divergent both at $0$ and at
$\infty.$ But splitting it into the two parts at the right, and interchanging the $d^4p$ with the $da_1da_2$ integrations
leaves a gaussian integral
\begin{equation*}
\int \exp (-\frac{c}{2}|p|^2)d^dp  = (2\pi/c)^{d/2}
\end{equation*}
which is convergent, but at the expense of getting $(a_1+a_2)^2$ in the denominator: The integral
\begin{equation*}
\int_0^\infty\int_0^\infty \frac{da_1da_2}{(a_1+a_2)^2}
\end{equation*}
 has a logarithmic singularity at $0$ and at $\infty.$ This can be
seen by blowing up the origin in $\R_{\ge 0}^2,$ and pulling back:
\begin{equation*}
\int_0^\infty \int_0^\infty \frac{db_1db_2}{b_1(1+b_2)^2}.
\end{equation*}
In other words, the trick with the parametric parameterization (called \emph{Schwinger trick} in \cite{BEK}), does
not get rid of any divergences. It just moves them into another, lower-dimensional space.  \\

Again it is useful to have a resolution of singularities in order to separate the various singularities and
divergences of a graph along irreducible components of a divisor with normal crossings. The most obvious and
efficient such resolution is given in \cite{BEK,BlKr}:\\

Let $\Gamma$ be core. For a subgraph $E(\gamma)\subseteq E(\Gamma),$ let
\begin{equation*}
L_\gamma = \cap_{e\in E(\gamma)} A_e=\{a_e=0,e\in E(\gamma)\},
\end{equation*}
 a linear subspace.
Set $\Lhi_{core}=\{L_\gamma: \gamma \mbox{ is a core subgraph of }\Gamma\},$ and
\begin{eqnarray*}
\Lhi_0 &=& \{\mbox{ minimal element of }\Lhi_{core}\}=\{0\}\\
\Lhi_{n+1} &=& \{\mbox{ minimal elements of }\Lhi_{core}\setminus \bigsqcup_{i=0}^n \Lhi_i\}
\end{eqnarray*}
This partition of $\Lhi_{core}$ is made in such a way that (see \cite[Proposition 3.1]{BlKr}) a sequence of blowups
\begin{equation}\label{eq:blowupseq}
\gamma: Z_S\rightarrow \ldots \rightarrow A_S
\end{equation}
is possible which starts by blowing up $\Lhi_0$ and then
successively the strict transforms of the elements of $\Lhi_1,\Lhi_2,\ldots$ This
ends up with $Z_S$ a manifold with corners. The map $\gamma$ is
of course defined not only as a map onto
$A_S=\R^{|E(\Gamma)|}_{\ge 0}$ but as a birational map $\gamma:
\mathcal{Z}_S\rightarrow \C^{|E(\Gamma)|},$ with
$\mathcal{Z}_S$ a smooth complex variety. The total
exceptional divisor $\mathcal{E}$ has normal crossings, 
and one component $\mathcal{E}_L$ for each $L\in \Lhi_{core}.$ (In the
language of section \ref{ss:ressing}, $\Lhi_{core}$ is the
building set). Moreover,
\begin{equation*}
\mathcal{E}_{L_1}\cap\ldots \cap \mathcal{E}_{L_k} \neq 0 \Longleftrightarrow \mbox{ the }L_i \mbox{ are totally ordered by inclusion.}
\end{equation*}
Since the coordinate divisor $\{a_e=0\mbox{ for some }e\in E(\Gamma)\}$ has already normal crossings by
definition, the purpose of these blowups is really only to pull out into codimension 1 all the intersections where
there are possibly singularities or divergences, and to separate the integrable singularities of the integrand from this set as much as possible. \\

Note that in the parametric situation where the domain of integration is the manifold with corners $\R^{|E(\Gamma)|}_{\ge 0},$
the blowups do not introduce an orientation issue on the real locus.\\

For the example graph $\Gamma_3$ of the previous sections (see (\ref{eq:examplegraph})),
\begin{equation*}
u_{\Gamma_3}^S = \frac{da_1\ldots da_6}{((a_1+a_2)((a_3+a_4)(a_5+a_6)+a_5a_6)+a_3a_4a_5+a_3a_4a_6+a_3a_5a_6)^{d/2}}
\end{equation*}
we examine the pullback
of $u_{\Gamma_3}^S$ onto $Z_S.$ There are various core subgraphs to consider, but it is easily seen, in complete analogy with
(\ref{eq:divarr}), that the divergences are located only at $L_{\Gamma_3},$ $L_{\gamma_2}$ and $L_{\gamma_1}$ where $\gamma_1$ is the
full subgraph with vertices 3 and 4, and $\gamma_2$ the full subgraph with vertices 2,3 and 4. In order to
see the divergences in $Z_S,$ it therefore suffices to look in a chart where $\mathcal{E}_{L_{\Gamma_3}},$ $\mathcal{E}_{L_{\gamma_2}}$ and $\mathcal{E}_{L_{\gamma_1}}$
intersect. In such a chart, given by coordinates
$b_1=a_1,$ $b_2=a_2/a_1,$ $b_3=a_3/a_1,$ $b_4=a_4/a_1,$ $b_5=a_5/a_3,$ $b_6=a_6/a_5,$
we have
\begin{equation}\label{eq:gamma3s}
\gamma^\ast u_{\Gamma_3}^S = \frac{db_1\ldots db_6}{b_1b_3b_5((1+b_2)((1+b_6)(1+b_4)+b_5b_6)+b_3(b_5b_6+b_4b_6+b_4))^{d/2}}
\end{equation}
Now we are in a very similar position as in the previous
section. If $\Gamma_p$ is a primitive
graph, then there is only the origin $0\in
A_S$ which needs to be blown up
in order to isolate the divergence. Since $u_{\Gamma_3}^S$ depends explicitly on $d$ in the exponent, let us
use $d$ as an analytic regulator.
One finds, using for example coordinates $b_1=a_1,$
$b_i=a_i/a_1,$ $i\neq 1,$ in a neighborhood of $d=4,$
\begin{equation*}
\gamma^\ast u_{\Gamma_p}^S(d) = \left(\frac{\delta_0(b_1)}{d-4}+\mbox{finite}\right)g_{\Gamma_p}
\end{equation*}
with $g_{\Gamma_p}\in L^1_{loc}.$ (If one wants even a regular
$g_{\Gamma_p}$ one needs to perform the remaining blowups in (\ref{eq:blowupseq}).)
Then we define
\begin{equation}\label{eq:paramres}
\operatorname{res}_S \Gamma_p = (\operatorname{res}_{d=4} \gamma^\ast u_{\Gamma_p}^S(d))
[\underline{1}]
=\int_{b_1=0,b_i\ge 0} g_{\Gamma_p} = \int_{\sigma} \frac{\Omega}{\Psi^2_{\Gamma_
p}}
\end{equation}
where $\sigma = \{a_i\ge 0\}\subset \P^{|E(\Gamma)|-1}(\R)$ and
$\Omega=\sum_{n=1}^{|E(\Gamma)|}(-1)^n a_n
da_1\wedge\ldots\wedge\widehat{da_n}\wedge\ldots\wedge
da_{|E(\Gamma)|}.$ The last integral at the right is a
projective integral, meaning that the $a_i$ are interpreted as
homogeneous coordinates of $\P^{|E(\Gamma)|-1}.$ By choosing
affine coordinates $b_i,$ one finds that it is identical with
the integral of $g_{\Gamma_p}$ over the
exceptional divisor
intersected with the total inverse image of $A_S.$ \\\\
Coming back to the non-primitive graph $\Gamma_3$ (see
(\ref{eq:gamma3s})) we find in complete analogy with section \ref{ss:ressing}, that
\begin{equation*}
u_{\Gamma_3}^S (d) = \sum_{n\ge -3}^\infty c_{n} (d-4)^n
\end{equation*}
in a neighborhood of $d=4,$ and
\begin{equation}\label{eq:stf2}
c_{-3}[\underline{1}_{A_S}] = (\operatorname{res}_S \Gamma_1)^3
\end{equation}
which is easily seen by sending $b_1,b_3,b_5$ to 0 in
(\ref{eq:gamma3s}):
$g_{\Gamma_3}|_{b_1=b_3=b_5=0}=g_{\Gamma_1}^{\otimes 3}.$\\

Similarly, one can translate the results of section
\ref{ss:ressing} and \cite{BBK} into this setting and obtain a renormalization
(extension of $u_\Gamma^S$) by removing the simple pole along
each component of the irreducible divisor. In section \ref{ss:blkren} a
different, motivic renormalization scheme for the parametric
representation will be studied, following \cite{BlKr}.
\subsection{Dyson-Schwinger equations}\label{ss:DSE}
Up to now we have only considered single Feynman graphs, with
internal edges interpreted as virtual particles, and parameters
such as the mass subjected to renormalization. Another approach
is to start with the full physical particles from the
beginning, that is, with the non-perturbative objects. Implicit
equations satisfied by the physical particles (full
propagators) and the physical interactions (full vertices) are
called \emph{Dyson-Schwinger equations.} The equations can
be imposed in a Hopf algebra of Feynman graphs \cite{BroadhurstKreimerDSE,Kreimer2,BK2,KreimerDSE,Yeats} and
turn into systems of integral equations when Feynman rules are
applied.\\

For general configurations of external momenta, Dyson-Schwinger
equations are extremely hard to solve. But if one sets all but
two external momenta to 0, a situation called zero-momentum
transfer (see (\ref{eq:phir})), then the problem simplifies considerably.\\

In \cite{Kreimeretude}, an example of a linear Dyson-Schwinger
equation is given which can be solved nonperturbatively by a
very simple Ansatz. More difficult non-linear Dyson-Schwinger
equations, and finally systems of Dyson-Schwinger equations as
above, are studied in \cite{KY,KY2,KY3,KY4}, see also
\cite{Yeats,KreimerNumbers,FoissySystems}.
\subsection{Remarks on minimal subtraction}\label{ss:misc}
I come back at this point to the difference between
what I call \emph{local} (section \ref{ss:ressing}) and \emph{global} (section \ref{ss:momentumspace})
\emph{minimal subtraction}, which, I think, is an important one. \\

I tried to emphasize in the exposition of the previous sections that
the key concepts of renormalization are largely independent of whether
momentum space, position space, or parametric space Feynman rules are
used. This is immediately seen in the Connes-Kreimer Hopf algebra
framework where a graph $\Gamma$ and some external
information $\sigma_\Gamma$ are sent directly to a Laurent series in $d-4.$
For this we don't get to see and don't need to know if the integral
has been computed in momentum, position, or parametric space. They all
produce the same number (or rather Laurent series), provided the same
regularization is chosen for all three of them.\\

In position space, where people traditionally like to work with distributions
as long as possible and integrate them against a test function only at the very
end (or even against the constant function $\underline{1},$ the adiabatic limit),
one is tempted to define the Feynman rules as a map into a space of distribution-valued
Laurent series, as we have done it in \cite{BBK}. But one has to be aware that
this space of distribution-valued Laurent series does not necessarily qualify as
a replacement for the ring $\mathcal{R}$ in section \ref{ss:momentumspace} if one looks
for a new Birkhoff decomposition. In general, many questions and misconceptions that I
have encountered in this area can be traced back to the decision at which moment one
integrates, and minimal subtraction seems to be a good example for this. \\

Let me now give a detailed comparison of what happens in local and global minimal
subtraction, respectively. Assume for example the massless graph in 4 dimensions
\begin{equation*}
\Gamma = \dunce
\end{equation*}
Clearly $\Gamma$ itself and the full subgraph $\gamma$ on the vertices 2 and 3
are logarithmically divergent. No matter which kind of Feynman rules we use,
assume there is a regularized Feynman distribution $u_\Gamma(\epsilon)$ varying
holomorphically in a punctured disk around $\epsilon=0,$ with a finite order pole
at $\epsilon=0.$ Assume after resolution of singularities that the regularized
Feynman distribution, pulled back onto the smooth model, has a simple pole supported on the
component $\mathcal{E}_\Gamma$ of the total exceptional divisor (for the superficial divergence),
and another on the component $\mathcal{E}_\gamma$ (for the subdivergence). Let $\mathcal{E}_\Gamma
= \{y_\Gamma=0\}$ and $\mathcal{E}_\gamma= \{y_\gamma=0\}$ in local coordinates $y_\Gamma,y_\gamma,y_3,\ldots,y_n.$
\begin{equation}\label{eq:ugammaeps}
u_\Gamma(\epsilon) = \left(\frac{\delta_0(y_\Gamma)}{\epsilon}+|y_\Gamma|_{fin}(\epsilon)\right)\left(\frac{\delta_0(y_\gamma)}{\epsilon}+|y_\gamma|_{fin}(\epsilon)\right) f_\Gamma(\epsilon)
\end{equation}
where $f_\Gamma$ is locally integrable and smooth in $y_\Gamma$ and $y_\gamma,$ such that in particular $f_\Gamma(\epsilon)$ is holomorphic in $\epsilon.$
There is accordingly a second order pole supported at $\mathcal{E}_\Gamma\cap \mathcal{E}_\gamma.$
We know from \cite{BBK}, as was also sketched in section \ref{ss:ressing}, that
the leading coefficient of this second order pole is a product of delta functions restricting it to $\mathcal{E}_\Gamma\cap
\mathcal{E}_\gamma$ times the residue of $\gamma$ times the residue of $\Gamma//\gamma.$ \\

Consequently, integrating $u_\Gamma(\epsilon)$ against a fixed function $\chi$ (for a first reading take $\chi=\underline{1}$
but in the massless case, one has to worry about infared divergences) provides a Laurent series
\begin{equation*}
u_\Gamma(\epsilon)[\chi] = a_{-2}\epsilon^{-2}+a_{-1}\epsilon^{-1}+a_0\epsilon^0+\ldots
\end{equation*}
Since $\gamma$ and $\Gamma//\gamma$ are primitive,
\begin{eqnarray*}
u_\gamma(\epsilon)[\chi] &=& b_{-1}\epsilon^{-1}+b_0\epsilon^0+b_1\epsilon^1+\ldots\\
u_{\Gamma//\gamma}(\epsilon)[\chi] &=& c_{-1}\epsilon^{-1}+c_0\epsilon^0+b_1\epsilon^1+\ldots
\end{eqnarray*}
We know from the previous remarks that $a_{-2}=\operatorname{res}(\gamma)\operatorname{res}(\Gamma//\gamma)=b_{-1}c_{-1}$
and similarly $a_{-1}=b_{-1}c_0+g$ where I don't want to specify $g.$ \\

Let me now compare local and global minimal subtraction at this example. Local minimal subtraction
is defined on distribution-valued Laurent series, but global minimal subtraction only on $\C$-valued
Laurent series. Therefore we need to integrate everything out before comparing. I start with local minimal subtraction (LMS).
In order to get from (\ref{eq:ugammaeps}) to
\begin{equation}\label{eq:ugammaren}
(u_\Gamma)_{R,\operatorname{LMS}}(\epsilon) = |y_\Gamma|_{fin}(\epsilon)|y_\gamma|_{fin}(\epsilon)f_\Gamma(\epsilon)
\end{equation}
one has to subtract three terms from (\ref{eq:ugammaeps}):
\begin{eqnarray*}
R_{\operatorname{LMS}}^\Gamma u_\Gamma(\epsilon) &=& \frac{\delta_0(y_\Gamma)}{\epsilon}\left(\frac{\delta_0(y_\gamma)}{\epsilon}+|y_\gamma|_{fin}(\epsilon)\right)f_\Gamma(\epsilon)\\
R_{\operatorname{LMS}}^{\gamma,\Gamma//\gamma} u_\Gamma(\epsilon) &=& \left(\frac{\delta_0(y_\Gamma)}{\epsilon}+|y_\Gamma|_{fin}(\epsilon)\right)\frac{\delta_0(y_\gamma)}{\epsilon}f_\Gamma(\epsilon)\\
-RR_{\operatorname{LMS}}^{\gamma,\Gamma//\gamma} u_\Gamma(\epsilon) &=& -\frac{\delta_0(y_\Gamma)}{\epsilon}\frac{\delta_0(y_\gamma)}{\epsilon}f_\Gamma(\epsilon)
\end{eqnarray*}
The first term cleans the pole supported on $\mathcal{E}_\Gamma,$ such that $u_\Gamma-R_{\operatorname{LMS}}^\Gamma u_\Gamma$ has only a simple pole supported on $\mathcal{E}_\gamma$ left.
On the other hand, $u_\Gamma-R_{\operatorname{LMS}}^{\gamma,\Gamma//\gamma} u_\Gamma$ has only a simple pole supported on $\mathcal{E}_\Gamma$ left, and the third term is a correction
term supported on $\mathcal{E}_\gamma\cap \mathcal{E}_\Gamma$ accounting for what has been subtracted twice.
In summary,
\begin{equation}\label{eq:lmsres}
(u_\Gamma)_{R,\operatorname{LMS}}(\epsilon)=u_\Gamma(\epsilon)-R_{\operatorname{LMS}}^\Gamma u_\Gamma(\epsilon)-R_{\operatorname{LMS}}^{\gamma,\Gamma//\gamma} u_\Gamma(\epsilon)+
RR_{\operatorname{LMS}}^{\gamma,\Gamma//\gamma} u_\Gamma(\epsilon)
\end{equation}
is the result of local minimal subtraction. \\

Let us now integrate out (\ref{eq:lmsres}).
\begin{eqnarray*}
u_\Gamma(\epsilon)[\chi] &=& a_{-2}\epsilon^{-2}+a_{-1}\epsilon^{-1}+a_0\epsilon^{0}+\ldots\\
R_{\operatorname{LMS}}^\Gamma u_\Gamma(\epsilon)[\chi] & =& a_{-2}\epsilon^{-2}+g \epsilon^{-1}+h\epsilon^0+\ldots\\
R_{\operatorname{LMS}}^{\gamma,\Gamma//\gamma} u_\Gamma(\epsilon)[\chi] &=& a_{-2}\epsilon^{-2}+b_{-1}c_0\epsilon^{-1}+b_{-1}c_{1}\epsilon^0+\ldots\\
-RR_{\operatorname{LMS}}^{\gamma,\Gamma//\gamma} u_\Gamma(\epsilon)[\chi] &=& a_{-2}\epsilon^{-2}
\end{eqnarray*}

These equations follow from (\ref{eq:ugammaeps}), and I don't want to specify $h.$ Consequently
\begin{equation*}
(u_\Gamma)_{R,\operatorname{LMS}}(\epsilon)[\chi]  =  a_0-b_{-1}c_1-h \mbox{ as }\epsilon\rightarrow 0.
\end{equation*}
In global minimal subtraction (GMS), where $R_{\operatorname{GMS}}=R$ as in (\ref{eq:r}), something different happens.
\begin{eqnarray*}
R_{\operatorname{GMS}}(u_\Gamma(\epsilon)[\chi]) &=& a_{-2}\epsilon^{-2}+a_{-1}\epsilon^{-1}\\
(R_{\operatorname{GMS}}u_\gamma(\epsilon)[\chi])u_{\Gamma//\gamma}(\epsilon)[\chi] &=& b_{-1}c_{-1}\epsilon^{-2}+b_{-1}c_0\epsilon^{-1}+b_{-1}c_1\epsilon^0+\ldots\\
-R_{\operatorname{GMS}}(R_{\operatorname{GMS}}u_\gamma(\epsilon)[\chi])u_{\Gamma//\gamma}(\epsilon)[\chi]) &=& b_{-1} c_{-1}\epsilon^{-2}+b_{-1}c_0\epsilon^{-1}
\end{eqnarray*}
The first subtraction $u_\Gamma[\chi]-R_{\operatorname{GMS}}(u_\Gamma[\chi])$ removes the poles everywhere, also the one supported on $\mathcal{E}_\gamma$ which
has nothing to do with the superficial divergence. The third and fourth term restore the locality of counterterms. We have
\begin{equation*}
(u_\Gamma)_{R,\operatorname{GMS}}(\epsilon)[\chi]  =  a_0-b_{-1}c_1 \mbox{ as }\epsilon \rightarrow 0.
\end{equation*}
In summary: Unless $h=0,$ local and global minimal subtraction differ by a finite renormalization. Moreover, although
there is a one-to-one-correspondence between terms to be subtracted in LMS and GMS, the values of those single terms do not agree.
It seems to me that GMS is a quite clever but somehow exceptional trick of defining the subtraction operator $R$ on $\C$-valued
Laurent series where all the geometric information (i.~e.~ where the pole is supported) has been forgotten.\\

In \cite{BBK} it is shown how to relate, for a general graph $\Gamma,$ the combinatorics of the total exceptional divisor of the resolution
of singularities to the Connes-Kreimer Hopf algebra of Feynman graphs, such that the example presented here is a special case of a more
general result. A similar analysis applies to other local renormalization prescriptions, called subtraction at fixed conditions in \cite{BBK},
as well.

\section{Motives and residues of Feynman graphs}\label{s:geometry}
\subsection{Motives, Hodge Realization and Periods}\label{ss:motives}
Much of the present interest in Feynman integrals is due to the
more or less obvious fact that there is something
\emph{motivic} about them. In order to understand and
appreciate this, one obviously needs to have an idea of what a
motive is. I am not an expert in this area and will not even
attempt to provide much background to the notion of motive. See
\cite{Andre} for an often cited introduction to the subject,
which I follow closely in the beginning of this section. \\

The theory of motives is a means to unify the various
cohomology theories known for algebraic varieties $X$ over a
number field $k.$ Such cohomology theories include the
algebraic de Rham and the Betti cohomology, but there are many
others. The algebraic de Rham cohomology $H^\bullet_{dR}(X)$ is
defined over the ground field $k$, and Betti cohomology
$H^\bullet_{B}(X;\Q)$ is the singular
cohomology of $X(\C)$ with rational coefficients.\\

A motive of a variety is supposed to be a piece of a universal
cohomology, such that all the usual cohomology theories
(functors from varieties to graded vector spaces) factor
through the category of motives. A particular cohomology theory
is then called a \emph{realization}. For example, the
combination of de Rham and Betti cohomology, giving rise to a
Hodge structure, is called \emph{Hodge realization}. \\

The theory of motives is not complete yet. Only for the
simplest kind of algebraic varieties, smooth projective ones, a
category of motives with the desired properties has been constructed.
These motives are called \emph{pure}. For general,
i.~e.~ singular or non-projective varieties, the theory is
conjectural in the sense that only a triangulated category
as a candidate for the derived category of the category of these motives,
called \emph{mixed motives} exists.\\

Let $X$ be a smooth variety over $\Q.$ Let
$H^\bullet_{dR}(X)$ denote the algebraic de Rham cohomology of
$X,$ a graded $\Q$-vector space, and $H^\bullet_{B}(X;\Q)$ the
rational Betti cohomology (singular cohomology of the complex
manifold $X(\C)$ with rational coefficients), a graded
$\Q$-vector space. A \emph{period} of $X$ is by definition a
matrix element of the \emph{comparison isomorphism}
(integration)
\begin{equation*}
H^\bullet_{dR}(X)\otimes_\Q \C \cong H_B^\bullet(X;\Q)\otimes_\Q \C
\end{equation*}
for a suitable choice of basis. A period is therefore in
particular an integral of an algebraic differential form over a
topological cycle on $X(\C).$ A standard example is the case of
an elliptic curve $X$ defined by the equation
$y^2=x(x-1)(x-\lambda),$ $\lambda\in\Q\setminus \{0,1\}.$ A basis element of $H^1_{dR}(X)$ is the
1-form $\omega=\frac{dx}{2y}$ and and a basis of the singular cohomology $H^1_{B}(X)$ is
given by the duals of two circles around the cut between $0$ and $1$ resp.~ the cut between $1$ and $\infty.$ Integrating $\omega$ against
these cycles gives the generators of the period lattice of $X.$ \\

Similarly, matrix elements of a comparison isomorphism between
\emph{relative} cohomologies of pairs $(X,A)$ are called
\emph{relative periods}. Many examples considered below will be
relative periods.
\subsection{Multiple zeta values, mixed Tate motives and the work of Belkale and Brosnan}\label{ss:bb}
Let $\Gamma$ be a primitive Feynman graph. I assume $d=4$ and
$m=0.$ Recall the graph polynomial
\begin{equation*}
\Psi_\Gamma = \sum_{T \operatorname{st} \operatorname{of} \Gamma} \prod_{e\not\in E(T)} a_e \in \Z[a_e: e\in E(\Gamma)]
\end{equation*}
from (\ref{eq:graphpoly}). The sum is over the spanning trees
of $\Gamma.$ Following \cite{BEK}, we have a closer look at the
parametric residue
\begin{equation*}
\operatorname{res}_S \Gamma = \int_\sigma\frac{\Omega}{\Psi^2_\Gamma}
\end{equation*}
introduced in (\ref{eq:paramres}). Let
$X_\Gamma=\{\Psi_\Gamma=0\}\subset \P^{|E(\Gamma)|-1}$ and
$CX_\Gamma=\{\Psi_\Gamma=0\}\subset \mathbb{A}^{|E(\Gamma)|}$
its affine cone. $X_\Gamma$ resp.~ $CX_\Gamma$ are called
\emph{projective} resp.~ \emph{affine graph hypersurface}. The
chain of integration is $\sigma=\{a_e\ge 0\}\subseteq
\P^{|E(\Gamma)|-1}(\R),$ and $\Omega=\sum (-1)^n a_n
da_1\wedge\ldots \wedge \widehat{d a_n} \wedge\ldots\wedge
da_{|E(\Gamma)|}.$ \\

The residue $\operatorname{res}_S \Gamma$ already looks like a
relative period, since $\sigma$ has its boundary contained in
the coordinate divisor $\Delta=\bigcup_{e\in
E(\Gamma)}\{a_e=0\},$ and the differential form
$\frac{\Omega}{\Psi^2_\Gamma}$ is algebraic (i.~e.~ regular) in
$\P^{|E(\Gamma)|-1}\setminus X_\Gamma.$ But in general
$X_\Gamma\cap \Delta$ is quite big, and
$\frac{\Omega}{\Psi_\Gamma^2}\not\in
H^{|E(\Gamma)|-1}_{dR}(\P^{|E(\Gamma)|-1}\setminus
X_\Gamma,\Delta\setminus (X_\Gamma\cap \Delta)).$ \\

The solution is of course to work in the blowup $\mathcal{Z}_S$
of section \ref{ss:parametric} where things are separated. Let
$\mathcal{P}_S$ be the variety obtained from
$\P^{|E(\Gamma)|-1}$ by regarding all elements of the $\Lhi_n$ $(n\ge 1)$ in section
\ref{ss:parametric} as subspaces of $\P^{|E(\Gamma)|-1}$ and
starting the blowup
sequence at $n=1$ instead of $n=0.$\\

In \cite{BEK,BlKr} it is shown that $\mathcal{P}_S$ has the
desired properties: the strict transform of $X_\Gamma$ does not
meet the strict transform of $\sigma.$ Like this,
$\operatorname{res}_S \Gamma$ is a relative period of the pair
\begin{equation*}
(\mathcal{P}_S\setminus Y_\Gamma,B\setminus (B\cap Y_\Gamma))
\end{equation*}
where $Y_\Gamma$ is the strict transform of $X_\Gamma,$ and $B$ the total transform of the coordinate divisor $\Delta.$ \\

We call $\operatorname{res}_S \Gamma$ a \emph{Feynman period}
of $\Gamma.$ \\

An empirical observation due to Broadhurst and
Kreimer \cite{BroadKreimer3,Broadhurst-Kreimer-96-2} was that
all Feynman periods computed so far are rational linear
combinations of multiple zeta values. \\

A \emph{multiple zeta value} of depth $k$ and weight
$s=s_1+\ldots+s_k$ is a real number defined as follows:
\begin{equation*}
\zeta(s_1,\ldots,s_k)=\sum_{1\le n_k<\ldots<n_1} \frac{1}{n_1^{s_1}\ldots n_k^{s_k}}
\end{equation*}
where $s_1\ge 2$ and $s_2,\ldots,s_k\ge 1.$ For $k=1$ one
obtains the values of the Riemann zeta function at integer
arguments $\ge 2,$ whence the name. \\

By an observation due to Euler and Kontsevich, multiple zeta
values can be written as iterated integrals
\begin{equation*}
\zeta(s_1,\ldots,s_k)=\int_{0<t_s<\ldots<t_1<1} w_{s_1}\wedge\ldots \wedge w_{s_k}
\end{equation*}
where
\begin{equation*}
w_{s}(t)=\left(\frac{dt}{t}\right)^{\wedge (s-1)}\wedge \frac{dt}{1-t}
\end{equation*}
and therefore qualify already as \emph{naive periods}, as defined in \cite{KZ}. \\

But in order to understand multiple zeta values as (relative)
periods of the cohomology of something, one needs to go one
step further and introduce the moduli space
$\mathcal{M}_{0,s+3}$ of genus $0$ curves with $s+3$ distinct
marked points, and its Deligne-Mumford compactification
$\overline{\mathcal{M}}_{0,s+3}.$

Indeed, starting from the iterated integral representation,
$\zeta(s_1,\ldots,s_k)$ can be shown to be a relative period of
a pair
\begin{equation*}
(\overline{\mathcal{M}}_{0,s+3}\setminus A,B\setminus (A\cap B))
\end{equation*}
with $A$ and $B$ suitable divisors which have no common
irreducible component. These pairs have \emph{mixed Tate} motives, a
special (and relatively simple and well-understood) kind of
mixed motives. This is a result of Goncharov and Manin
\cite{GoMa}. Brown showed that conversely every such relative
period of $\overline{\mathcal{M}}_{0,s+3}$ is a rational linear combination of multiple zeta values \cite{Brown4}.\\

Let us now come back to the Feynman periods. Even up to now,
not a single example of a Feynman period is known which is not
a rational linear combination of multiple zeta values.
Moreover, these multiple zeta values do not arise randomly, but
there are already certain
patterns visible. For examples of such patterns, see \cite{BroadKreimer3,Broadhurst-Kreimer-96-2,BEK,Schnetzcensus}. \\

Motivated by an (informal) conjecture of Kontsevich
\cite{Kontsevichconj}, Belkale and Brosnan investigated the
motives associated to Feynman graph hypersurfaces. Kontsevich's
conjecture did not state directly that all Feynman periods be
multiple zeta values, but that the function
\begin{equation*}
q \mapsto |CX_\Gamma(\mathbb{F}_q)|
\end{equation*}
be a polynomial in $q$ for all $\Gamma.$ Using another
conjecture about motives, a non-polynomial counting function
for the number of points of $CX_\Gamma$ over $\mathbb{F}_q$
would imply that $CX_\Gamma$ has a period which is \emph{not}
in the $\Q$-span of multiple zeta values. For example, an
elliptic curve is known to have a non-polynomial point counting function.\\

Belkale and Brosnan came to the surprising result that
Kontsevich's conjecture is false \cite{BB}, and that Feynman
graph hypersurfaces have the most general motives one can think
of.

\subsection{Matroids and Mn\"ev's theorem}

One key idea in Belkale's and Brosnan's proof was to study more
general schemes defined by matroids:
\begin{dfn} Let $E$ be a finite set and $I\subseteq 2^E.$ The pair $M=(E,I)$ is called \emph{matroid} if
\begin{enumerate}
\item $\emptyset\in I,$
\item $A_1\subseteq A_2,$ $A_2\in I$ $\Longrightarrow$ $A_1\in I,$
\item $A_1,A_2\in I,$ $|A_2|>|A_1|$ $\Longrightarrow$ there is an $x\in A_2\setminus A_2\cap A_1$ such that
$A_1\cup \{x\}\in I.$
\end{enumerate}
The number $\operatorname{rk} M=\operatorname{max}_{A\in I}|A|$ is called \emph{rank of} $M.$
\end{dfn}
The subsets $A\in I$ where $|A|$ is maximal are called \emph{bases} of $M.$
The literature usually names two standard examples for matroids:
\begin{enumerate}
\item $M=(E,I)$ where $E$ is a finite set of vectors in some $k^r,$ $I$ the set of linearly independent
subsets of $E.$ Clearly $\operatorname{rk} M\le r.$
\item $M=(E,I)$ where $E$ is the set of edges of a graph
    and $I$ the set of subgraphs (each determined by a
    subset of edges) without cycles. Clearly
    $\operatorname{rk} M = |V(\Gamma)|-\operatorname{rk}
    H_0(\Gamma;\Z).$
\end{enumerate}
We have already seen in section \ref{ss:rules} how these
examples are related (in fact, the second is a special case of
the first): If $\Gamma$ is a graph, for each $e\in E(\Gamma)$
there is a linear form $e^\vee j_\Gamma$ on
$\R^{|V(\Gamma)|}/H_0(\Gamma;\R),$ and such linear forms
$e_1^\vee j_\Gamma,\ldots,e_n^\vee j_\Gamma$ are pairwise
linearly independent if and only if the graph with edges
$\{e_1,\ldots,e_n\}$ has no cycles.\\

Let us return to the general case. A matroid is equivalently
characterized by a rank function on $2^E$ as follows:

\begin{dfn} A map $r: 2^E \rightarrow \N$ is called \emph{rank function} if
\begin{enumerate}
\item $r(A)\le |A|,$
\item $A_1\subseteq A_2$ $\Longrightarrow$ $r(A_1)\le r(A_2),$
\item $r(A_1\cup A_2)+r(A_1\cap A_2)\le r(A_1)+r(A_2).$
\end{enumerate}
\end{dfn}
\begin{pro} Let $M=(E,I)$ be a matroid. Then the map
\begin{equation*}
r: A\mapsto \operatorname{rk}(A,\{B\in I, B\subseteq A\})
\end{equation*}
is a rank function. Conversely, let $E$ be a finite set and $r$
a rank function for it. Then $M=(E,r)=(E,I)$ where
$I=\{A\subseteq E, r(A)=|A|\}$ is a matroid. \hfill $\Box$
\end{pro}

We have seen how linearly independent subsets of vectors in a
vector space give rise to a matroid. On the other hand one may
ask if every matroid is obtained this way:
\begin{dfn} Let $k$ be a field. A matroid $M=(E,r)$ is called \emph{realizable over} $k$ is there is an $r\in\N$ and
a map $f:E\rightarrow k^r$ with $\dim \operatorname{span}
f(A)=r(A)$ for all $A\in 2^E.$ Such a map is called \emph{
representation of $M$}.
\end{dfn}
There are matroids which are representable only over certain fields, for example the Fano matroid.\\

The space $X(M,s)$ of all representations of $M$ in $k^s$ (a
subvariety of $\mathbb{A}^{s|E|}$ defined over $k$) is called
\emph{representation space} of $M.$ It is a fundamental
question how
general these realization spaces are. An answer is given by Mn\"ev's Universality Theorem.\\

Mn\"ev's Universality Theorem was originally proved by
Mn\"ev in a context of \emph{oriented matroids} and their
representations over the ordered field of real numbers. Without giving a
precise definition, an oriented matroid keeps not only track of
whether or not certain subsets of vectors are linearly
dependent but also about the sign of determinants: Roughly an
oriented matroid is specified by a list of partitions of $E$
indicating which vectors in $E$ may be separated by linear
hyperplanes in $\R^n.$ Again the representation space of an
oriented matroid is the space of vector configurations which
leaves this list of partitions invariant. The original, quite
difficult, version of the theorem is then
\begin{thm}[Mn\"ev, oriented version] For every primary semi-algebraic set $X$ in $\R^r$ defined over $\Z$ there
is an oriented matroid whose realization space is
stably equivalent to $X.$
\end{thm}
Here a primary semi-algebraic set defined over $\Z$ is a set
given by polynomial equations and \emph{sharp} polynomial
inequalities $<,>$ with integer coefficients, (such as
$x_1^2+x_2^2>2,$ $x_2x_1^3=1$), and stable equivalence means
roughly a sort of homotopy equivalence preserving certain
arithmetic properties. The proof in Mn\"ev's thesis
\cite{Mnevthesis,Mnevrev} is quite
intricate, and there is a simplified proof in \cite{RGrevisited,BB}
which I follow here.\\

The simpler version that we need is obtained by
replacing primary semi-algebraic sets by affine schemes of finite type over $\operatorname{Spec} \Z,$
oriented matroids by matroids, and stable equivalent by isomorphic with an open subscheme in a product with $\mathbb{A}^N.$
Just like the affine representation space, there is a projective representation space
\begin{eqnarray*}
 \hat X(M,s) &=& \{ f: E \hookrightarrow \P^{s-1}: \\
 && \operatorname{dim} \operatorname{span} f(A) = r(A)-1 \mbox{ for all } A\in 2^E\}
\end{eqnarray*}

\begin{thm}[Mn\"ev, un-oriented version]\label{thm:mnev2}
Let $X$ be an affine scheme of finite type over $\operatorname{Spec} \Z.$ Then there is a
matroid $M$ of rank 3, $N\in\N$ and an open $U\subseteq X\times \mathbb{A}^N$ projecting surjectively onto $X$ such that
\begin{equation*}
U\cong \hat X(M,3)/PGL_3.
\end{equation*} \hfill $\Box$
\end{thm}

This is the version in Lafforgue's book \cite{Lafforgue}. I am grateful
to A.~ Usnich for showing me this reference. See also \cite{BoStBook} for the independently obtained version of Sturmfels.\\

Suppose $X$ is defined by $f_+-f_-=0$ where $f_+$ and $f_-$ are
polynomials with positive coefficients. The $f_\pm$ can be
successively decomposed into more elementary expressions
involving only one addition or one multiplication at a time, at
the expense of introducing many more variables. The proof of
\tref{thm:mnev2} uses then the fact that once $x_1$ and $x_2$
are fixed on a projective line, $x_1+x_2$ and $x_1x_2$ etc.~
can be determined by linear dependence conditions in the
projective plane (this is why the rank of $M$ is only 3). The
difficulties left are to relate different projective scales and
to avoid unwanted dependencies.\\

Like this any affine scheme over $\operatorname{Spec} \Z$ is related to the
representation space of a (huge) rank 3 matroid. Belkale and
Brosnan use a slightly different version of Mn\"ev's theorem and
then show (a lot of work that I just skip) how this
representation space is connected to
the graph hypersurfaces $CX_\Gamma.$ \\

Let me now state the main result of \cite{BB}: Let
$\operatorname{GeoMot}^+$ be the abelian group with generators
isomorphism classes $[X]$ of schemes $X$ of finite type over
$\Z$ modulo the relation
\begin{equation*}
[X]=[X\setminus V]+[V]
\end{equation*}
if $V$ is a closed subscheme of $X.$ Endowed with the cartesian
product $[X][Y]=[X\times Y],$ $\operatorname{GeoMot}^+$ becomes
a ring with unit $[\operatorname{Spec} \Z ].$ Let
$L=[\mathbb{A}^1]$ be the Tate motive, and $S$ the saturated
multiplicative subset of $\Z[L]$ generated by $L^n-L$ for
$n>1.$ Let
$\operatorname{GeoMot}=S^{-1}\operatorname{GeoMot}^+,$ and
$\operatorname{Graphs}$ the $S^{-1}\Z[L]$-submodule of
$\operatorname{GeoMot}$ generated by the $[CX_\Gamma],$ where
$\Gamma$ are Feynman graphs.
\begin{thm}[Belkale, Brosnan]
$\operatorname{Graphs} = \operatorname{GeoMot}.$ \hfill $\Box$
\end{thm}
It is clear that point-counting $q\mapsto |X(\mathbb{F}_q)|$
factors through $\operatorname{GeoMot}.$ Therefore Kontsevich's
conjecture is false. Also it is known \cite[Section 15]{BB}
that the mixed Tate property can be detected in
$\operatorname{GeoMot}.$ Therefore it follows that not all
$X_\Gamma$ are mixed Tate, and (using another conjecture) that
not all periods of all $X_\Gamma$ are rational linear
combinations of multiple zeta values. \\

On the other hand, not all periods of all $X_\Gamma$ are
Feynman-periods in the sense defined in section \ref{ss:bb}.

\subsection{The work of Bloch, Esnault and Kreimer}\label{ss:bek}
A finer study of motives of certain Feynman graph hypersurfaces
is carried out in the second part of \cite{BEK}: For the so
called \emph{wheels with $n$ spokes},
\begin{equation*}
WS_n = \wheel
\end{equation*}
one has
\begin{thm}[Bloch, Esnault, Kreimer]\label{thm:bek}
\begin{equation*}
H_c^{2n-1}(\P^{2n-1}\setminus X_{WS_n})\cong \Q(-2),
H^{2n-1}(\P^{2n-1}\setminus X_{WS_n}) \cong \Q(-2n+3)
\end{equation*}
and $H^{2n-1}_{dR}(\P^{2n-1}\setminus X_{WS_n})$ is generated
by $\Omega/\Psi^2_{WS_n}.$ \hfill $\Box$
\end{thm}

It had been known before
\cite{BroadKreimer3,Broadhurst-Kreimer-96-2} that
\begin{equation*}
\operatorname{res}_S WS_n \in \zeta(2n-3) \Q^\times
\end{equation*}
and \tref{thm:bek} partially confirms that an extension
\begin{equation*}
0 \rightarrow \Q(2n-3)\rightarrow E \rightarrow \Q(0)\rightarrow 0
\end{equation*}
is responsible for this (see \cite[Section
9]{BEK},\cite[Section 9]{BlochJapan}).

\subsection{The work of Bloch and Kreimer on renormalization}\label{ss:blkren}
Let us return to renormalization. Within the parametric Feynman
rules, Bloch and Kreimer \cite{BlKr} show how to understand
renormalized non-primitive integrals using periods of a
\emph{limiting mixed Hodge structure.} \\

Limiting mixed Hodge structures arise in a situation where
there is a family of Hodge structures varying over a base
space, in this case a punctured disk $D^\ast$ (For zero
momentum transfer Feynman graphs this one-dimensional base
space is sufficient). In contrast to section
\ref{ss:parametric}, the parameter $t\in D^\ast$ does not alter
the exponent of the differential form, but is
rather some sort of cut-off for the chain of integration. \\

It follows from our discussion in \ref{ss:parametric} that the
projective integral
\begin{equation*}
\int_\sigma \frac{\Omega}{\Psi^2_\Gamma}
\end{equation*}
is not convergent unless $\Gamma$ is primitive (This is the
reason why $\operatorname{res}_S\Gamma$ is defined only for
primitive integrals): there are poles along the exceptional
divisors $\mathcal{E}_{L_\gamma}$ corresponding to divergent
subgraphs $\gamma.$ In other words, $\int_\sigma
\frac{\Omega}{\Psi^2_\Gamma}$ is not a period. But by varying
the coordinate divisor $\Delta_t$ (and the simplex $\sigma_t$
sitting inside $\Delta_t)$ with $t\in D^\ast,$ one has a
\emph{family} of mixed Hodge structures, and for all $t\neq 0$
the period $\int_{\sigma_t} \frac{\Omega}{\Psi^2_\Gamma}$ is defined.\\

Bloch and Kreimer describe how to express the monodromy
operation on (relative) homology, in particular on $\sigma_t,$
in terms of suitable tubes around the strata of the exceptional
divisor of $\mathcal{Z}_S.$ Winding around such a tube picks up
the residue along the stratum (see section
\ref{ss:parametric}). Since the monodromy is quasi-unipotent,
its logarithm gives a (graph-independent) nilpotent matrix $N$
such that
\begin{equation}\label{eq:bk}
\int_{\sigma_t}\frac{\Omega}{\Psi^2_\Gamma} = \mbox{first row of }\exp(N\log t/2\pi i)(a_1,\ldots,a_r)^t
\end{equation}
up to a multi-valued analytic function vanishing at $t=0,$ with
$a_1,\ldots,a_r$ periods of a limiting mixed Hodge structure \cite{BlKr}.\\

When there is only one non-zero external momentum, say $P,$ the
relation between the regularization (\ref{eq:bk}) and the
renormalized integral (\ref{eq:phir}), where the second graph
polynomial must be taken into account, is easy to see.
Therefore (\ref{eq:bk}) also tells about the coefficients
$p_n(\Gamma)$ of the renormalized integral (\ref{eq:phir}), and
one observes in the monodromy representation the same
combinatorial objects (nested sets, the Connes-Kreimer
coproduct) that have guaranteed locality of counterterms in
section \ref{s:regren}.

\subsection{Final remarks}\label{ss:outlook}

Let me finish this second part of the paper by just mentioning very briefly some
other results that have been obtained in this area. \\

The Belkale-Brosnan theorem does not provide a specific
counterexample graph to Kontsevich's conjecture (it does
provide a counterexample matroid). See
\cite{Doryntalk,Schnetz} for recent developments in this
direction.

The methods of \cite{BEK} have been extended in \cite{Dzmitry}
to other graphs than the wheels with spokes.
Regularization and renormalization in the parametric
representation is also discussed in
\cite{BognerWeinzierl1,BognerWeinzierl2,Marcolli}.

The relation between Feynman periods and multiple zeta values
as periods of the moduli space of stable genus 0 curves is
studied much further in \cite{Brown2,Brown3}. Finally the reader may be interested in \cite{BlochX,AM1,AM2,AM4,Patterson}
for a further study of graph hypersurfaces.

\bibliographystyle{alpha}
\nocite{Usnich,Bergbauerthesis,BB2,BK2,CMSP}
\bibliography{lit}
\end{document}